\documentclass[modern]{aastex63}

\usepackage{epsfig}
\usepackage{epstopdf}
\usepackage{longtable}
\usepackage{amsmath}
\usepackage{multirow}
\usepackage{hyperref}
\usepackage{upgreek}
\usepackage{fancyhdr}
\usepackage{booktabs}

\begin{document}

\title{Ground Vibrational State S\MakeLowercase{i}O Emission in the VLA BA\MakeLowercase{a}DE Survey}

\author{V. Dike}
\affil{Department of Physics and Astronomy, University of California Los Angeles}
\author{M. R. Morris}
\affil{Department of Physics and Astronomy, University of California Los Angeles}
\author{R. M. Rich}
\affil{Department of Physics and Astronomy, University of California Los Angeles}
\author{M. O. Lewis}
\affil{Department of Physics and Astronomy, University of New Mexico}
\affil{National Radio Astronomy Observatory}
\author{L. H. Quiroga-Nu{\~n}ez}
\altaffiliation{L. H. Quiroga-Nu{\~n}ez is a Jansky Fellow of the National Radio Astronomy Observatory.}
\affil{National Radio Astronomy Observatory}
\affil{Department of Physics and Astronomy, University of New Mexico}
\author{M. C. Stroh}
\affil{Center for Interdisciplinary Exploration and Research in Astrophysics and Department of Physics and Astronomy, Northwestern University}
\author{A. C. Trapp}
\affil{Department of Physics and Astronomy, University of California Los Angeles}
\author{M. J. Claussen}
\affil{National Radio Astronomy Observatory}

\begin{abstract}
	
Using a subsample of the Bulge Asymmetries and Dynamical Evolution (BAaDE) survey of stellar SiO masers, we explore the prevalence and characteristics of $^{28}$SiO $J=1-0, v=0$ emission. We identify 90 detections of maser, thermal, or composite $^{28}$SiO $J=1-0, v=0$ emission out of approximately 13,000 candidate spectra from the NSF's Karl G. Jansky Very Large Array (VLA). We find that the detected sources are likely asymptotic giant branch (AGB) stars belonging to a bright, foreground Milky Way stellar disk population. For the 32 sources showing thermal components, we extract values for outflow velocity by fitting thermal line profiles. We find a range of circumstellar envelope expansion velocities, and compare to previously recorded OH and CO expansion velocities. This preliminary survey is already the largest study of stellar ground-vibrational-state SiO masers to date, and will be expanded to include the entire VLA BAaDE dataset when data reduction for the 18,988 target sources is completed.   

\end{abstract}

\section{Introduction}\label{sec:intro}

SiO emission, thought to be produced above the atmospheres of AGB stars in material that has been levitated by stellar pulsations, is a powerful tool for probing the conditions of the AGB circumstellar environment. SiO masers have long been observed in the excited vibrational states ($v \ge 1$) since their discovery in the Orion nebula by \cite{SnyBuhl1974}, but, with a few exceptions, they have been found in the circumstellar envelopes of evolved, mass-losing stars. 

Originally, the emission found in the ground vibrational state was observed to be thermal (in this paper, “thermal'' means unaffected by maser amplification, rather than “thermalized''). The first surveys of stellar SiO emission focused on the $J=2-1, v=0$ thermal line (\citealt{Dickinson1978}, \citealt{MorrisSiO1979}). Thermal lines are useful because their line width gives a measure of the local velocity of the stellar outflow, and their line intensity provides a measure of how much of the gas-phase SiO in the extended envelope has survived grain formation \citep{MorrisAlcock1977}. 

Circumstellar masing from the $J=1-0, v=0$ state was predicted by the radiative pumping model of \cite{Kwan1974}. A tentative detection of maser emission from the ground vibrational state ($v=0$) of $^{28}$SiO in NML Cyg was reported by \cite{Dickinson1978}. \cite{chandler1995} observed the $^{28}$SiO $J=1-0$ ground state transition in the IRc2 star-forming region in Orion, characterizing the emission profile as a combination of maser and thermal. \cite{jewell1991} found the first stellar SiO emission from the $J=1-0, v=0$ state of $^{28}$SiO and recorded 12 sources described as thermal but noted that some had spikes that could be maser emission. \cite{Boboltz2004} reported ground-state maser emission in at least 5 sources, and, more recently, \cite{devicente2016} reported combined profiles of thermal and maser emission in a total of 28 evolved stars. These profiles have a narrow maser spike above a broad base of thermal emission.

SiO masers are known to occur closer to the stellar photosphere than other circumstellar masers such as OH and H$_{2}$O. \cite{Linc1995} showed with a Very Long Baseline Array study of the AGB star VX Sgr that SiO masers ($J=1-0, v=1$) are produced just above the photosphere at 1.3 times the stellar radius (see also \citealt{Diamond2003}), but interferometric observations of the spatial distribution of the ground-state maser line emission indicate that it is produced farther out in the envelope \citep{Boboltz2004}. This is consistent with the infrared pumping model (including both collisions and radiation) of \cite{Herpin2000}, which predicts the $J=1-0, v=0$ emission to be farther from the star in regions of lower SiO density. 

The pumping mechanism for SiO maser emission, including $v=0$ maser emission, is not agreed upon. A population inversion from radiative pumping by stellar photons is suggested by the radial alignment of the polarized emission of SiO $J=1-0, v=1$ and, $v=2$ emission \citep{Desmurs2000}. However, a model of collisional pumping is supported by observations of SiO $J=1-0, v=1$ and $v=2$ lines because they arise in the same region of space around the star \citep{miyoshi1994}. Predictions for $J=1-0, v=0$ emission from either model are scarce, likely because of the similarly scarce detections. 

In this paper, we report a subsample of stars exhibiting SiO $J=1-0, v=0$ emission at 43424 MHz from the Bulge Asymmetries and Dynamical Evolution (BAaDE) dataset\footnote{\url{http://www.phys.unm.edu/~baade/}}. BAaDE is the largest survey of stellar SiO maser sources, so we are able to study the largest  sample of  SiO $J=1-0, v=0$ emission to date. Our line profiles have a variety of shapes: narrow maser lines, broad thermal lines, and a composite of the two. The survey and our subsample are discussed in section \ref{sec:obs}. In section \ref{sec:methods}, we discuss our method for modeling and fitting our thermal and composite lines. In section \ref{sec:lines}, we show our detections and the results of our line fitting, including our stellar expansion velocities. Section \ref{sec:pop} puts our sources in context by examining the characteristics of the subsample and how they compare to other sources in the full BAaDE survey. Section \ref{sec:pop} also relates our calculated line widths with these population characteristics. We summarize and discuss future work in section \ref{sec:disc}.

\section{Observations and Sample Selection}\label{sec:obs}

The BAaDE survey uses the the NSF's Karl G. Jansky Very Large Array (VLA) and Atacama Large Millimeter Array (ALMA) to survey SiO masers in the Milky Way, with the main goal of constraining kinematics of bulge stars by using the masers to measure stellar line-of-sight velocity. The radio frequencies of SiO masers ensure that there is minimum attenuation by dust in the Galactic plane. Each source was observed for 40-50 seconds. The BAaDE VLA frequency range, 42.4 to 43.4 GHz, encompasses the SiO $J=1-0, v=0,1,2,3$ transitions, with a frequency resolution of 1.7 km s$^{-1}$. BAaDE observations also include the 86 GHz SiO $J=2-1, v=0,1,2$ emission in the ALMA dataset---a southern sample of stars complementary to that observed with the VLA---as presented by \cite{Michael2019}. 

The BAaDE sample was selected using the criterion that the stars fall within the \textit{iiia} region of the Midcourse Space Experiment (MSX) color-color plane, which is most likely to contain stars with oxygen-rich circumstellar envelopes, and therefore SiO masers (\citealt{msx2009}; see also \citealt{Megan2020}). The total sample includes 28,062 stars, 18,988 of which were observed with the VLA. For a more detailed description of the survey, see \cite{Lorant2017} and \cite{Lorant2018}. For early results showing that the disk and bulge populations of the BAaDE survey are distinct kinematically, see \cite{Adam2018}. The velocity of SiO masers is the same as the stellar velocity (see \cite{jewell1991}), which makes them a useful probe of bulge kinematics. Statistics of the vibrational SiO line transitions in a subsample of BAaDE sources are explored in \cite{Michael2018}, in which the authors use maser line strength ratios from different vibrational states as a diagnostic of the density of the circumstellar envelope. Moreover, bolometric characteristics as well as the spatial distribution of the BAaDE sample within the solar neighborhood are presented in \cite{LHQN2020}.


In this paper, we use a subset of the VLA dataset to study the $^{28}$SiO $J=1-0, v=0$ line at 43424 MHz. In order to characterize the SiO $v=0$ lines from the BAaDE survey, representative lines were first selected by visually inspecting the BAaDE VLA data that had been calibrated and processed by August 2018, corresponding to roughly 2/3 of the final VLA survey size. Based on the characteristics of the  $v=0$ thermal and maser lines seen at 43424 MHz, which are always much weaker than the vibrationally excited transitions, an automatic line-detection algorithm was devised. An interval of 58 MHz (400 km s$^{-1}$) was searched on each side of the 43423.853 MHz rest frequency. First, this interval was searched for channels of flux density exceeding 5 times the RMS noise. Any secondary peaks are ignored and the highest amplitude channel assumed to be the line center. To search for broad lines specifically, the same frequency interval was first searched for any channels above the 5-$\sigma$ threshold, and these channels  removed from the spectrum. Then the same frequency interval was again searched using a 5-channel window with a detection threshold for the mean amplitude of the channels in the window to be seven times the RMS noise divided by the square root of the number of channels in the window. The search using the 5-channel window was applied on a channel-by-channel basis. The same velocity range was then similarly searched with a 10-channel window, but we counted a detection with either window size as long as the signal-to-noise ratio of the integrated intensity was above seven, as defined above. Most broad-line sources were found with both window sizes.

Both the narrow and broad line-finding algorithms were run on approximately 13,000 BAaDE spectra. The detections were then visually inspected to reject false-positives, leaving a total of 90 significant detections. These detections are shown in Figures \ref{mfig}, \ref{fittfig}, and \ref{fitcfig} in the Appendix. The galactic coordinates $l$ and $b$ of our detected $v=0$ sources, corresponding 2 Micron All Sky Survey (2MASS) names, and any corresponding OH or variable star name are listed in Table \ref{coordnames}. We cross-matched with the database\footnote{\url{maserdb.net}} collected by \cite{maserdb} using a $1\arcsec$  matching radius to the 2MASS coordinates if known (MSX coordinates otherwise) and note whether there are any previously recorded SiO observations.

\newpage
\startlongtable
\begin{deluxetable}{cDDcccc}
	\centerwidetable
	\tablecaption{Galactocentric coordinates (degrees) and alternative names for $v=0$ sources. The rightmost column lists references to any previous SiO maser observations recorded in the database of \cite{maserdb}. \label{coordnames}}
	\tablehead{\colhead{Source Name} & \multicolumn2c{$l$} & \multicolumn2c{$b$} & \colhead{2MASS name}  & \colhead{OH name} & \colhead{V* name} & \colhead{Ref.} }
	\decimals
	\startdata
	\cutinhead{Maser-only} %
	{ad3a-00230}	&	$-$4.41	&	$-$2.98	&	  J17464903$-$3415068	&	OH355.588$-$02.978	&		&		\\	%
	{ad3a-00531}	&	$-$7.37	&	3.01	&	  J17145167$-$3325546	&	OH352.625+03.014	&	RW Sco	&	\tablenotemark{a} \tablenotemark{b} \tablenotemark{c} \tablenotemark{d}	\\	%
	{ad3a-00562}	&	$-$4.84	&	$-$0.60	&	  J17355922$-$3321388	&	OH355.156$-$00.597	&		&		\\	%
	{ad3a-00582}	&	$-$8.87	&	5.21	&	  J17021034$-$3319285	&		&		&		\\	%
	{ad3a-01244}	&	$-$1.93	&	$-$1.74	&	  J17475111$-$3129244	&		&		&		\\	%
	{ad3a-01758}	&	$-$1.77	&	0.12	&	  J17405413$-$3022380	&	OH358.235+00.115	&		&	\tablenotemark{c}	\\	%
	{ad3a-01815}	&	$-$1.64	&	0.09	&		&		&		&		\\	%
	{ad3a-01975}	&	$-$1.56	&	0.44	&	  J17400664$-$3002019	&		&		&		\\	%
	{ad3a-02502}	&	0.17	&	$-$1.31	&	  J17511077$-$2928102	&		&		&		\\	%
	{ad3a-04289}	&	$-$1.62	&	3.79	&	  J17270541$-$2815299	&		&		&	\tablenotemark{d}	\\	%
	{ad3a-04410}	&	$-$0.05	&	1.59	&	  J17391937$-$2808357	&	OH359.946+01.593	&		&		\\	%
	{ad3a-04942}	&	1.63	&	0.20	&	  J17484093$-$2725587	&		&		&		\\	%
	{ad3a-04944}	&	1.58	&	0.29	&	  J17481382$-$2725523	&		&		&	\tablenotemark{e}	\\	%
	{ad3a-05408}	&	0.25	&	4.23	&	  J17300394$-$2628029	&		&	V2311 Oph	&		\\	%
	{ad3a-05488}	&	4.89	&	$-$3.12	&	  J18085198$-$2616210	&	OH004.887$-$03.121	&		&	\tablenotemark{f}	\\	%
	{ad3a-05782}	&	3.10	&	1.68	&	  J17462494$-$2525010	&		&		&		\\	%
	{ad3a-06568}	&	4.06	&	4.05	&	  J17394605$-$2321241	&		&	V545 Oph	&		\\	%
	{ad3a-07151}	&	5.60	&	3.39	&	  J17454035$-$2223291	&		&		&		\\	%
	{ad3a-08004}	&	10.41	&	0.31	&	  J18072713$-$1946392	&		&		&		\\	%
	{ad3a-08141}	&	9.89	&	2.37	&	  J17584623$-$1913286	&		&		&		\\	%
	{ad3a-08264}	&	12.77	&	$-$1.94	&	  J18203699$-$1847096	&		&		&		\\	%
	{ad3a-08428}	&	12.21	&	0.38	&	  J18105305$-$1810076	&		&		&		\\	%
	{ad3a-08663}	&	13.57	&	$-$0.30	&	  J18160768$-$1717563	&		&		&		\\	%
	{ad3a-09257}	&	16.12	&	$-$0.28	&	  J18210454$-$1502362	&		&		&		\\	%
	{ad3a-10192}	&	20.76	&	$-$1.47	&	  J18341639$-$1129291	&		&		&		\\	%
	{ad3a-10275}	&	20.16	&	0.25	&	  J18265500$-$1113473	&		&		&		\\	%
	{ad3a-10590}	&	19.73	&	2.42	&	  J18181792$-$1035450	&		&		&		\\	%
	{ad3a-11089}	&	23.33	&	$-$1.74	&	  J18400377$-$0920051	&		&		&		\\	%
	{ad3a-11091}	&	24.23	&	$-$3.45	&	  J18475335$-$0919102	&		&		&		\\	%
	{ad3a-11358}	&	24.13	&	$-$0.58	&	  J18372231$-$0805358	&		&		&		\\	%
	{ad3a-11369}	&	23.97	&	$-$0.16	&	  J18353454$-$0802402	&	OH023.970$-$00.164	&		&		\\	%
	{ad3a-11532}	&	24.21	&	0.63	&	  J18330976$-$0727583	&	OH024.208+00.633	&		&		\\	%
	{ad3a-12084}	&	26.12	&	$-$0.66	&	  J18412025$-$0621581	&		&		&	\tablenotemark{g}	\\	%
	{ad3a-12356}	&	27.29	&	$-$0.85	&	  J18440895$-$0524198	&		&		&		\\	%
	{ad3a-12687}	&	28.28	&	$-$0.15	&		&	OH28.28 $-$0.15	&		&		\\	%
	{ad3a-12759}	&	28.72	&	$-$0.58	&	  J18454842$-$0400462	&	OH28.7$-$0.6	&	V0439 Sct	&		\\	%
	{ad3a-13063}	&	29.4	&	$-$0.51	&	  J18464926$-$0322330	&		&		&		\\	%
	{ad3a-13192}	&	29.94	&	$-$0.87	&	  J18490520$-$0303361	&		&		&		\\	%
	{ad3a-13249}	&	30.59	&	$-$1.62	&	  J18525523$-$0249236	&		&		&		\\	%
	{ad3a-13536}	&	$-$147.41	&	$-$3.73	&	  J06373133$-$0123430	&		&	SY Mon	&	\tablenotemark{b} \tablenotemark{c} \tablenotemark{d}	\\	%
	{ad3a-13641}	&	33.37	&	1.10	&	  J18482151$-$0051232	&		&		&		\\	%
	{ad3a-14145}	&	33.80	&	$-$0.24	&	  J18535249+0039313	&		&		&	\tablenotemark{h}	\\	%
	{ad3a-15065}	&	40.13	&	2.42	&	  J18555696+0730302	&	OH40.1+2.4	&		&		\\	%
	{ad3a-15253}	&	44.80	&	$-$2.31	&	  J19213663+0927567	&	OH44.8$-$2.3	&		&		\\	%
	{ad3a-15268}	&	44.27	&	$-$0.99	&	  J19155397+0936556	&		&	V1839 Aql	&		\\	%
	{ad3a-17013}	&	71.74	&	$-$2.94	&	  J20233617+3215108	&		&		&		\\	%
	{ad3a-17084}	&	72.92	&	$-$2.14	&	  J20234087+3340405	&		&		&		\\	%
	{ad3a-17227}	&	83.66	&	$-$10.17	&	  J21272829+3641557	&		&	V1906 Cyg	&		\\	%
	{ae3a-00115}	&	0.52	&	$-$0.21	&		&		&		&		\\	%
	{ae3a-00116}	&	0.47	&	$-$0.11	&	  J17470929$-$2835290	&		&		&		\\	%
	{ae3a-00249}	&	23.60	&	$-$0.69	&	  J18364758$-$0836451	&		&		&		\\	%
	{ae3a-00272}	&	25.23	&	0.18	&	  J18364204$-$0645549	&		&		&		\\	%
	{ae3a-03912}	&	71.91	&	$-$34.26	&		&		&		&		\\	%
	{ce3a-00005}	&	$-$4.94	&	$-$0.70	&	  J17361012$-$3329407	&		&		&		\\	%
	{ce3a-00142}	&	20.18	&	$-$1.70	&	  J18340225$-$1206501	&		&		&		\\	%
	{ce3a-00147}	&	22.00	&	0.07	&	  J18310242$-$0941134	&	OH021.996+00.072	&		&		\\	%
	{ce3a-00163}	&	26.89	&	0.20	&	  J18394138$-$0516558	&		&		&		\\	%
	{ce3a-00181}	&	30.89	&	0.20	&	  J18465890$-$0143428	&	OH030.885+00.203	&		&		\\	%
	\cutinhead{Thermal$-$only} %
	{ad3a-00073}	&	$-$5.41	&	$-$2.29	&	  J17412754$-$3444239	&		&		&		\\	%
	{ad3a-01076}	&	$-$3.35	&	$-$0.32	&	  J17384049$-$3157182	&	OH356.64$-$0.32	&		&	\tablenotemark{b} \tablenotemark{c}	\\	%
	{ad3a-04662}	&	1.47	&	$-$0.37	&	  J17503125$-$2751565	&		&		&		\\	%
	{ad3a-06634}	&	5.91	&	1.23	&	  J17542613$-$2314097	&		&	V0774 Sgr	&		\\	%
	{ad3a-08239}	&	11.72	&	$-$0.18	&	  J18115787$-$1852125	&		&	V4724 Sgr	&		\\	%
	{ad3a-09805}	&	19.82	&	$-$2.85	&	  J18373291$-$1258051	&		&		&		\\	%
	{ad3a-09843}	&	18.27	&	0.36	&	  J18225398$-$1251080	&		&		&		\\	%
	{ad3a-11288}	&	25.09	&	$-$3.02	&	  J18475454$-$0821287	&	OH025.086$-$03.022	&		&	\tablenotemark{c} \tablenotemark{f}	\\	%
	{ad3a-11547}	&	24.23	&	0.64	&	  J18331043$-$0726170	&		&		&		\\	%
	{ad3a-11702}	&	25.33	&	$-$0.81	&	  J18402409$-$0708030	&		&		&		\\	%
	{ad3a-13264}	&	29.42	&	0.81	&	  J18420843$-$0245154	&		&		&	\tablenotemark{c}	\\	%
	{ae3a-00012}	&	$-$4.19	&	0.28	&	  J17340928$-$3220006	&		&		&		\\	%
	\cutinhead{Composite} %
	{ad3a-01844}	&	$-$1.84	&	0.49	&	  J17391493$-$3014243	&	OH358.162+00.490	&	V1019 Sco	&		\\	%
	{ad3a-01922}	&	$-$1.33	&	$-$0.04	&	  J17423501$-$3005419	&	OH358.667$-$00.044	&		&	\tablenotemark{d}	\\	%
	{ad3a-04112}	&	1.10	&	$-$0.83	&	  J17512677$-$2825371	&	OH001.095$-$00.832	&		&	\tablenotemark{d}	\\	%
	{ad3a-04509}	&	2.93	&	$-$3.23	&	  J18050223$-$2801541	&		&	V1804 Sgr	&	\tablenotemark{b} \tablenotemark{c}	\\	%
	{ad3a-05214}	&	2.58	&	$-$0.43	&	  J17531884$-$2656374	&		&		&		\\	%
	{ad3a-05396}	&	4.97	&	$-$3.74	&	  J18112756$-$2629324	&	OH5.0 $-$3.8	&		&		\\	%
	{ad3a-06531}	&	6.66	&	$-$0.45	&	  J18022487$-$2325161	&		&		&		\\	%
	{ad3a-07589}	&	8.93	&	$-$0.01	&	  J18053549$-$2113422	&	OH008.933$-$00.014	&		&	\tablenotemark{b} \tablenotemark{c}	\\	%
	{ad3a-07741}	&	6.52	&	4.98	&	  J17415458$-$2047055	&		&		&	\tablenotemark{f}	\\	%
	{ad3a-07850}	&	9.54	&	0.77	&		&		&		&		\\	%
	{ad3a-09606}	&	17.55	&	$-$0.13	&	  J18231790$-$1342473	&	OH017.551$-$00.126	&		&	\tablenotemark{b} \tablenotemark{c} \tablenotemark{d}	\\	%
	{ad3a-12266}	&	26.58	&	$-$0.11	&	  J18401349$-$0542113	&		&		&		\\	%
	{ad3a-12641}	&	27.93	&	0.24	&	  J18412530$-$0420320	&		&	V0646 Sct	&	\tablenotemark{d}	\\	%
	{ad3a-13434}	&	30.82	&	$-$0.16	&	  J18480882$-$0156540	&	OH030.823$-$00.156	&		&	\tablenotemark{d}	\\	%
	{ad3a-13789}	&	32.06	&	2.56	&	  J18395534$-$0018426	&		&		&		\\	%
	{ae3a-00123}	&	0.55	&	$-$0.06	&	  J17470898$-$2829561	&		&		&		\\	%
	{ce3a-00110}	&	8.34	&	$-$1.00	&	  J18080404$-$2213266	&	OH008.344$-$01.002	&	VX Sgr	&	\tablenotemark{a} \tablenotemark{b} \tablenotemark{c} \tablenotemark{d}	\\	%
	{ce3a-00127}	&	14.17	&	$-$0.06	&	  J18162603$-$1639563	&		&	V5102 Sgr	&	\tablenotemark{b} \tablenotemark{c}	\\	%
	{ce3a-00172}	&	28.45	&	0.13	&	  J18424805$-$0355396	&		&		&		\\	%
	{ce3a-00180}	&	30.94	&	$-$0.16	&	  J18482194$-$0150363	&		&		&		\\	%
	\enddata
	\tablerefs{    	\tablenotetext{a}{\cite{cho96}}
		\tablenotetext{b}{\cite{kim10}}
		\tablenotetext{c}{\cite{kim14}}
		\tablenotetext{d}{\cite{maserdb}}
		\tablenotetext{e}{\cite{fuj06}}
		\tablenotetext{f}{\cite{deg07b}}
		\tablenotetext{g}{\cite{deg04b}}
		\tablenotetext{h}{\cite{deg10a}} }  
	
\end{deluxetable}

\newpage
\section{Methods}\label{sec:methods}

We fit our thermal SiO lines to the model for thermal emission profiles from spherically symmetric envelopes expanding at constant velocity developed in \cite{MarkScan1985} (after \citealt{sobolev1960} and \citealt{Mark1975}):

	\begin{equation}	
	T_b(v_z) = T_0 \dfrac{\bigg(1-\Big[\dfrac{v_z-{v_c}}{V}\Big]^2\bigg)\bigg(1-\exp\Big[-\alpha\Big(1-\Big[\dfrac{v_z-{v_c}}{V}\Big]^2\Big)^{-1}\Big]\bigg)}{1-\exp(-\alpha)} , 
	\label{marquation}
	\end{equation}
 
 \vspace{5 mm} 
\noindent in which $T_b$ is the measured brightness temperature, $v_z$ is the velocity of each channel, $T_0$ is the brightness temperature at the height of the center of the profile, $v_c$ is the velocity at the center of the profile (and hence the line-of-sight velocity of the star), $V$ is the expansion velocity of the stellar envelope, and $\alpha$ is an opacity term parameterizing the shape of the profile which becomes parabolic (concave down) in the optically thick limit and flat-topped in the optically thin limit. This model assumes that the profile is symmetric and that the source is unresolved in most cases.

To perform this fit, we first converted our data from flux density as a function of observing frequency to brightness temperature as a function of velocity. We used the following version of the Rayleigh-Jeans equation to convert from flux density $S_{\nu}$ to brightness temperature:

	\begin{equation}	
 T_b = \dfrac{2S_{\nu}c^2}{\pi k \nu^2 \theta^2} = 5.05\times 10^5 S_{\nu}  \mathrm{(Jy)},
		\label{rjquation}
	\end{equation}

\vspace{5 mm} 
\noindent where $c$ is the speed of light, $k$ the Boltzmann constant, $\nu$ the observing frequency of each channel, and $\theta$ is the half-power beam width. Here we use the estimate of $\theta$ to be 0.043 arcseconds for the VLA Q-band in A configuration, which we use because we are assuming the sources are unresolved, instead of using the value for the C or D configurations, which were the actual observing configurations. We converted the frequency of each channel to velocity using the Doppler equation, $v_z = c(\nu - \nu_0)\nu_0^{-1}$, in which the rest frequency of the $J= 1-0, v=0$ line (43423.853 MHz) is $\nu_0$. 

The four-parameter Equation \ref{marquation} was fit to the profiles using the nonlinear least-squares fitter CURVEFIT\footnote{\url{https://www.harrisgeospatial.com/docs/curvefit.html}} in IDL 8.7. Sources with combined profiles---those with a maser spike above a plateau of thermal emission---were fit after zero-weighting the maser emission, i.e. not allowing those points that are part of the maser spike to influence the fit. Sources showing only thermal emission were fit without modification. Velocities were converted to the local standard of rest at the end of the process.

\section{Results}\label{sec:results}
\subsection{Model Fits}\label{sec:lines}

We identified 90 total sources of ground-state emission. Of these, 58 have only narrow maser emission (Figure \ref{mfig}). We successfully fit Equation \ref{marquation} to the remaining 32 sources exhibiting thermal emission, 20 of which also show maser emission. Figures \ref{fittfig} and \ref{fitcfig} in the Appendix present the spectra of the thermal and composite sources, respectively.

In Table \ref{fitparams} we report the values of our model fits of parameters $T_0$, $\alpha$, and $v_c$. The errors reported are the $1\sigma$ standard deviations from CURVEFIT. In the cases where there is maser emission that is zero-weighted for the fit, we caution about the value of our opacity parameter $\alpha$ for the shape of the line profile because a significant portion of the center of the line profile is missing. A complete study of opacity would require using more than one rotational transition, and brightness temperature varies with the phase of the stellar oscillation for our sources, so these parameters are reported for completeness but not discussed further in this paper. For profiles with both maser and thermal components, our values of $v_c$ coincide with the velocity of the maser peak in most cases, which indicates that the profiles are symmetric as we assumed and that the maser velocity is a good indicator of the stellar velocity. The outliers are ad3a-01844, ad3a-06531, ad3a09606, and ce3a-00127 (see Figure \ref{fitcfig}). The interpretation of the profiles is ambiguous as to whether the maser is truly offset or there is noise at the edge of the thermal pedestal. 

The most relevant parameter obtained from our fitting process is  $V$, the expansion velocity of the circumstellar envelope, which is shown in Table \ref{Vtable}, along with previously derived values of expansion velocity from OH and CO observations. In the model we adopt from \cite{MarkScan1985}, $V$ is half the full width at zero intensity of the thermal line in km s$^{-1}$ because the outflow is assumed to be symmetric about the velocity of the star itself. Out of 8 OH expansion velocities matched to those in \cite{OH1997} and \cite{OH2001}, 3 are consistent within 1$\sigma$. Furthermore, our SiO outflow velocities coincide with the CO outflow velocities of the 4 sources matched to \cite{CO1998}.

Sources with only a thermal component are shown in Figure \ref{fittfig}, and sources with both thermal and maser components are shown in Figure \ref{fitcfig}.
\newpage
\startlongtable
\begin{deluxetable}{ccccc}
	\tablecaption{Three of the four parameters from our fit, using the thermal emission model of \cite{MarkScan1985}. The fourth parameter, $V$, is shown in Table \ref{Vtable}. $T_0$ is brightness temperature, $\alpha$ is the opacity parameter, and $v_c$ is the local standard of rest velocity of the star, which has an intrinsic error of about 1 km s$^{-1}$. The fourth column shows signal-to-noise ratio of the integrated intensity. The number in parentheses is the 1$\sigma$ error in the last digit. \label{fitparams}}
	\tablehead{\colhead{Source Name} & \colhead{$T_0$} & \colhead{$\alpha$} & \colhead{$v_c$}  & \colhead{SNR} \\
		&  \colhead{$10^3$K}  & \colhead{} & \colhead{km s$^{-1}$} & \colhead{} }
	\decimals
	\startdata
		\cutinhead{Thermal-only}
{ad3a-00073}	&	28(2)	&	0.9(5)	&	-111(1)	&	9.0	\\	%
{ad3a-01076}	&	28(4)	&	1.7(8)	&	6(1)	&	7.3	\\	%
{ad3a-04662}	&	54(2)	&	4(3)	&	11.1(3)	&	7.8	\\	%
{ad3a-06634}	&	36(3)	&	0.6(4)	&	-2.2(8)	&	17.3	\\	%
{ad3a-08239}	&	32(4)	&	0.6(6)	&	-37(1)	&	8.1	\\	%
{ad3a-09805}	&	23(5)	&	$1^{+7}_{-1}$	&	65.3(8)	&	7.2	\\	%
{ad3a-09843}	&	19.(3)	&	32(15)	&	37(1)	&	7.6	\\	%
{ad3a-11288}	&	24(2)	&	1.1(8)	&	8.0(8)	&	8.3	\\	%
{ad3a-11547}	&	43(3)	&	13(6)	&	47.8(7)	&	19.9	\\	%
{ad3a-11702}	&	19.1(6)	&	0.20(6)	&	78.6(2)	&	9.6	\\	%
{ad3a-13264}	&	25(2)	&	0.7(5)	&	56.0(8)	&	17.0	\\	%
{ae3a-00012}	&	21(3)	&	1.1(9)	&	-99(2)	&	7.4	\\	%
		\cutinhead{Composite}		
{ad3a-01844}	&	35(5)	&	3(1)	&	4(1)	&	10.3	\\	%
{ad3a-01922}	&	34.3(8)	&	2.6(7)	&	-3.1(3)	&	13.8	\\	%
{ad3a-04112}	&	50(2)	&	0.5(1)	&	10.2(2)	&	8.8	\\	%
{ad3a-04509}	&	62(11)	&	9(6)	&	30.1(6)	&	9.1	\\	%
{ad3a-05214}	&	33(5)	&	2(1)	&	-1(2)	&	12.1	\\	%
{ad3a-05396}	&	21(3)	&	1.0(9)	&	28(1)	&	7.7	\\	%
{ad3a-06531}	&	12(9)	&	0.6(3)	&	12.0(6)	&	7.5	\\	%
{ad3a-07589}	&	29(1)	&	1.0(2)	&	13.5(6)	&	12.3	\\	%
{ad3a-07741}	&	33(4)	&	30(28)	&	32.1(7)	&	12.1	\\	%
{ad3a-07850}	&	121(15)	&	4(1)	&	13.4(5)	&	13.6	\\	%
{ad3a-09606}	&	80(2)	&	0.32(5)	&	44.3(1)	&	15.1	\\	%
{ad3a-12266}	&	20.2(7)	&	0.15(6)	&	-41.1(2)	&	8.3	\\	%
{ad3a-12641}	&	96(6)	&	11(3)	&	52.1(6)	&	18.0	\\	%
{ad3a-13434}	&	30(2)	&	1.0(5)	&	93.2(6)	&	11.1	\\	%
{ad3a-13789}	&	31.3(9)	&	0.8(1)	&	-44.1(3)	&	10.9	\\	%
{ae3a-00123}	&	73(9)	&	9(4)	&	-24.8(9)	&	13.6	\\	%
{ce3a-00110}	&	143(2)	&	1.8(2)	&	7.4(2)	&	16.9	\\	%
{ce3a-00127}	&	34(2)	&	0.6(2)	&	49.9(8)	&	15.0	\\	%
{ce3a-00172}	&	12(2)	&	0.6(6)	&	82(1)	&	10.1	\\	%
{ce3a-00180}	&	33(4)	&	0.8(7)	&	91.7(9)	&	23.5	\\	%
		\enddata
		
	\end{deluxetable}

\startlongtable
\begin{deluxetable}{cccc}
	\tablecaption{Expansion velocity parameter $V$ from fitting our profiles along with expansion velocities from previous observations of OH (\citealt{OH1997} and \citealt{OH2001}) and CO \citep{CO1998}. \label{Vtable}}
	\tablehead{\colhead{Source Name} & \colhead{$V$} & \colhead{$V_{\rm OH}$} & \colhead{$V_{\rm CO}$} }
	\decimals
	\startdata
	\cutinhead{Thermal-only}
	 	{ad3a-00073}	&	28(2)	&	--	&	--	\\	%
		{ad3a-01076}	&	18(2)	&	--	&	--	\\	%
		{ad3a-04662}	&	11.4(3)	&	--	&	--	\\	%
		{ad3a-06634}	&	29(1)	&	--	&	30.6 (1-0), 30.4 (2-1)	\\	%
		{ad3a-08239}	&	12(1)	&	--	&	--	\\	%
		{ad3a-09805}	&	6(6)	&	--	&	--	\\	%
		{ad3a-09843}	&	19(2)	&	--	&	--	\\	%
		{ad3a-11288}	&	15.1(9)	&	13.6	&		\\	%
		{ad3a-11547}	&	28(1)	&	--	&	27.6 (2-1)	\\	%
		{ad3a-11702}	&	16.6(2)	&	--	&	--	\\	%
		{ad3a-13264}	&	25(1)	&	--	&	--	\\	%
		{ae3a-00012}	&	30(3)	&	--	&	--	\\	%
	\cutinhead{composite}
		{ad3a-01844}	&	20(2)	&	19.7	&	--	\\	%
		{ad3a-01922}	&	26.1(4)	&	21.9	&	--	\\	%
		{ad3a-04112}	&	9.9(2)	&	19.7	&	--	\\	%
		{ad3a-04509}	&	14(1)	&	--	&	--	\\	%
		{ad3a-05214}	&	28(2)	&	--	&	--	\\	%
		{ad3a-05396}	&	20(1)	&	--	&	--	\\	%
		{ad3a-06531}	&	13.7(7)	&	--	&	--	\\	%
		{ad3a-07589}	&	29.5(6)	&	4.6	&		\\	%
		{ad3a-07741}	&	15(1)	&	--	&	16.0 (1-0), 18.0 (2-1)	\\	%
		{ad3a-07850}	&	15(1)	&	--	&	--	\\	%
		{ad3a-09606}	&	14.1(2)	&	14.8	&	--	\\	%
		{ad3a-12266}	&	18.2(2)	&	--	&	--	\\	%
		{ad3a-12641}	&	30(1)	&	--	&	--	\\	%
		{ad3a-13434}	&	15.2(7)	&	20.5	&	--	\\	%
		{ad3a-13789}	&	21.5(4)	&	--	&	--	\\	%
		{ae3a-00123}	&	24(2)	&	--	&	--	\\	%
		{ce3a-00110}	&	24.3(2)	&	19.3	&	22 (2-1)	\\	%
		{ce3a-00127}	&	30(1)	&	--	&	--	\\	%
		{ce3a-00172}	&	18(1)	&	--	&	--	\\	%
		{ce3a-00180}	&	32(2)	&	--	&	--	\\	%
	\enddata
		
\end{deluxetable}

\subsection{Population Characteristics}\label{sec:pop}

\begin{figure}
	\figurenum{1}
	\makebox[\textwidth][c]{\includegraphics[width=1.5\textwidth]{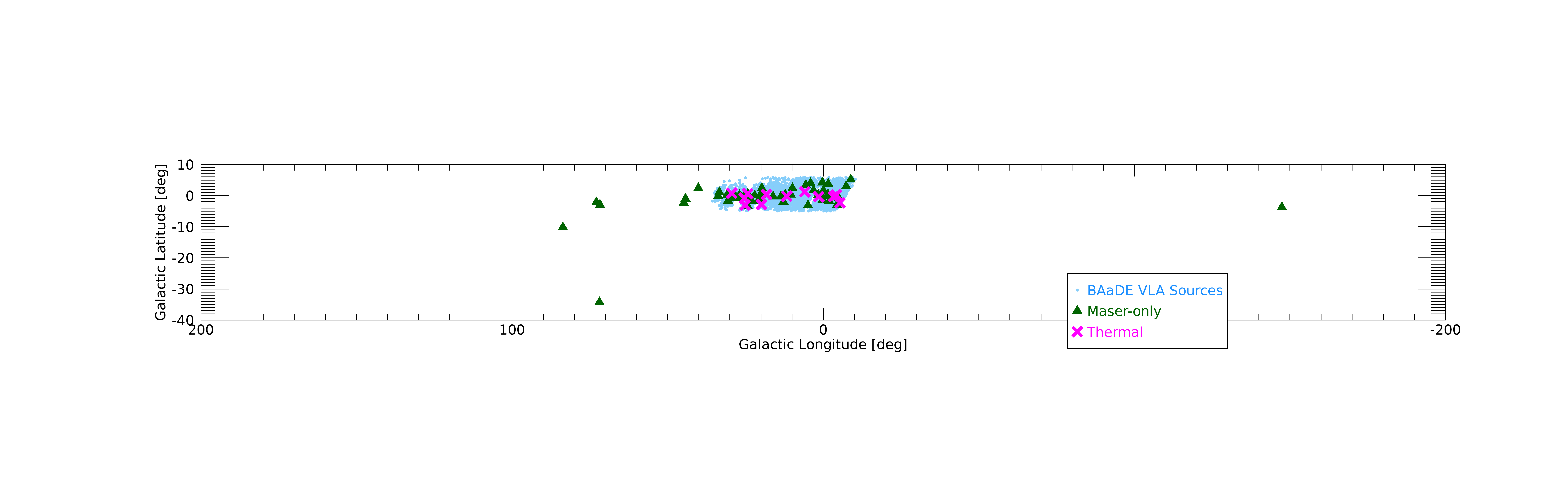}}%
	\caption{Distribution of maser and thermal sources on the sky, along with the current BAaDE VLA survey detections. Outlying maser sources are from observing runs not yet fully reduced. \label{posfig}}
\end{figure}

\begin{figure}
	\figurenum{2}
	\makebox[\textwidth][c]{\includegraphics[width=1.5\textwidth]{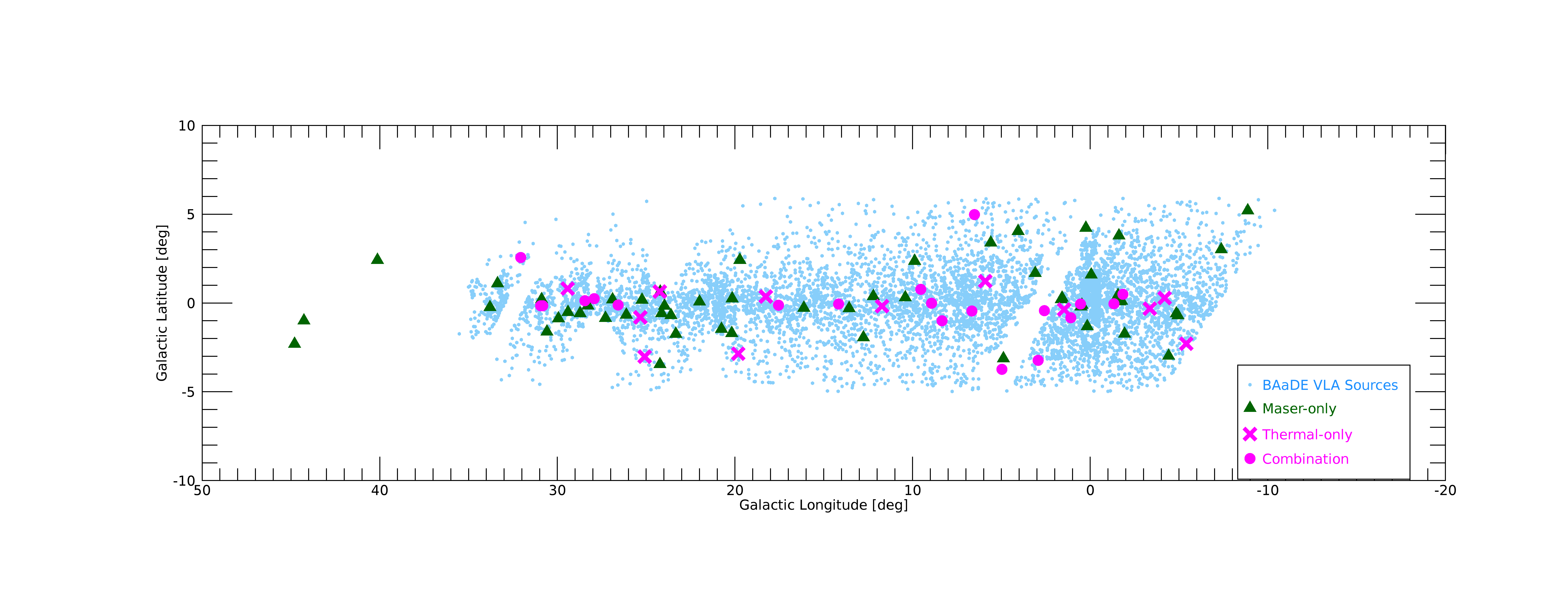}}%
	\caption{An expanded view of the inner part of Figure \ref{posfig}, where most of our sources are located. \label{zoomposfig}}
\end{figure}

\begin{figure}
	\figurenum{3}
	\fig{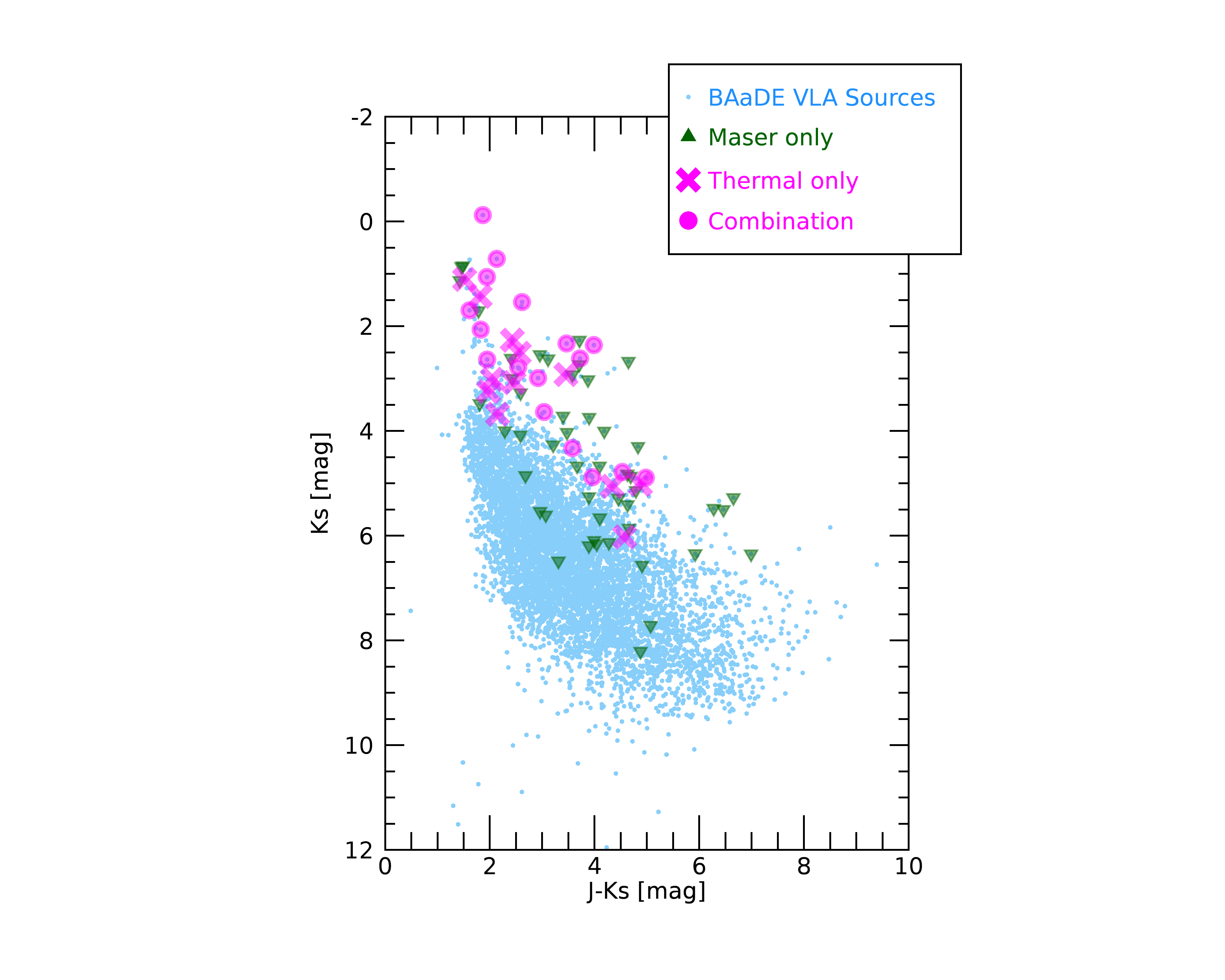}{1\textwidth}{}
	\caption{2MASS J-Ks color against 2MASS Ks magnitude. The maser emission can be seen in sources with dimmer Ks magnitude than the thermal sources. This is likely a simple selection effect because the maser emission in our sample is typically brighter than the thermal emission.\label{cmdfig}}
\end{figure}

\begin{figure}
	\figurenum{4}
	\fig{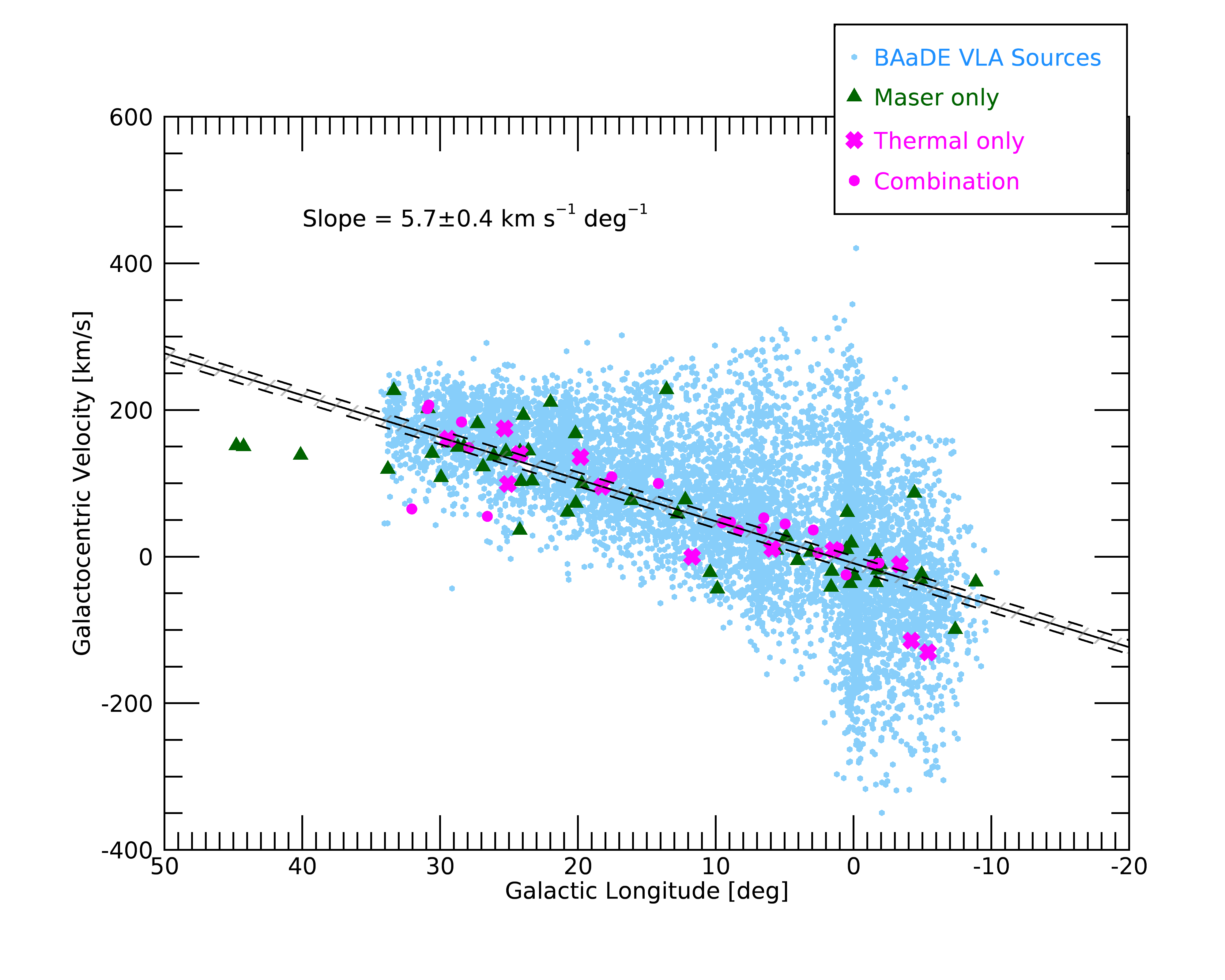}{.5\textwidth}{}
	\caption{Velocity distribution with longitude of our maser and thermal sources with the BAaDE VLA survey sample for reference. The line is fit only to the ground-state maser and thermal sources that we analyze in this work, excluding an outlying maser farther than 50$\degr$ from the Galactic center. The slope of the line describes the collective kinematics of our ground-state emission population. \label{kinfig}}
\end{figure}

We find evidence that our  $J=1-0, v=0$ sources belong to the Milky Way stellar disk. The locations of our detections, along with all the BAaDE pointings to which our line-finding algorithm has been applied, are shown in Figure \ref{posfig}. There is no latitudinal clustering of our sources in a specific region of the sky, which is better seen in Figure \ref{zoomposfig}, which shows the locus of the majority of our sources toward the central region of the Galaxy. As shown in Figure \ref{cmdfig}, our sources tend to be brighter than the main locus of the BAaDE population. This characteristic is consistent with that of the `kinematically cold' population in the work of \cite{Adam2018}, in which stars in the BAaDE sample with Ks magnitude $<$ 5.5 are found to have a lower collective velocity dispersion and a smaller velocity gradient because of Galactic rotation than stars with Ks $>$ 5.5. This is supported by examining the galactocentric velocities of our population as a function of Galactic longitude in Figure \ref{kinfig}. We also plot the slope of the rotation trend in longitude, fitting a line in velocity-longitude space to our sources. The rotation slope we measure, 5.73 km s$^{-1}$ deg$^{-1}$, is consistent with the model of foreground disk stars in \cite{Adam2018} as well as with the rotation slope of the `kinematically cold' BAaDE population of approximately 5.5 km s$^{-1}$ deg$^{-1}$ identified therein to be foreground stars in the thick disk. The 'kinematically hot' population has a slope of ~9 km s$^{-1}$ deg$^{-1}$ and primarily consists of members of the Galactic bulge. The fact that our sample of stars with v=0 emission detected can be ascribed to a foreground disk population could result from a selection effect, given that $v=0$ emission is weak compared to the $J=1-0, v=1,2,3$ masers so we see only nearby sources. 

We were able to find 9 distance estimates using matches to the \textit{Gaia} DR2 sample (\citealt{Gaia2016}, \citealt{Gaia2018}), assuming no priors when the relative parallax error was less than 20\% \citep{distance2015}, shown in Table \ref{distable}. The distances, almost entirely within a kiloparsec, are consistent with the hypothesis that our sample of stars is in the foreground. From this matching, we also find that three sources are classified by \textit{Gaia} as Mira variables: \mbox{ad3a-00531}, \mbox{ad3a-09606}, and \mbox{ce3a-00127}. The crossmatch between \textit{Gaia} and BAaDE and the subsequent extraction of distances is the subject of a separate study (Quiroga-Nu{\~n}ez et al., submitted).

\begin{deluxetable}{ccc}
	\tablecaption{Distance estimates from \textit{Gaia} parallax measurements. \label{distable}}
	\tablehead{\colhead{Source Name} & \colhead{Distance} & \colhead{Distance Error}  \\
		 & \colhead{(pc)} & \colhead{(pc)} }
	\decimals
	\startdata
	ad3a-00230 &   478 &       74 \\
	ad3a-00531 &   514 &       54 \\
	ad3a-08239 &   683 &      110 \\
	ad3a-09606 &   981 &      190 \\
	ad3a-10192 &   737 &      144 \\
	ad3a-12641 &  1078 &      170 \\
	ad3a-13536 &   539 &       70 \\
	ad3a-13789 &   548 &       99 \\
	ce3a-00127 &   828 &      131 \\
	\enddata
			
	\end{deluxetable}

We found no correlation with the apparent MSX-D magnitude (15 $\upmu$m). This indicates that the brightness of a source does not correlate with the expansion velocity.

All of our sources also express at least one other maser from higher vibrational states, including the $v=1,2$, or 3 states of $^{28}$SiO, and in a majority of cases, the $^{29}$SiO $v=0$ maser. We find the fraction of $^{29}$SiO $v=0$, $^{29}$SiO $v=1$, and $^{30}$SiO $v=0$ masers present in our sample to be higher than that of the greater BAaDE survey. For our sources, 74\% have a $^{29}$SiO $v=0$ line, 2\% have a $^{29}$SiO $v=1$ line, and 38\% have a $^{30}$SiO $v=0$ line, compared to 15\%, 0.5\%, and 2\% detection rate, respectively, in BAaDE sources that have a $^{28}$SiO $v=1$ line. Fluxes for isotopologue lines in the BAaDE survey will be published in Sjouwerman et al. (2020, in prep). Our higher fractions can be explained by the fact that these lines are also weak, so sources that are strong enough to display $^{28}$SiO $v=0$ emission are often also strong enough to detect any isotopologue emission that may be present. However, we cannot rule out a pumping mechanism may be favorable to both $J=1-0, v=0$ emission and these isotopologue lines.

\section{Discussion and Conclusions}\label{sec:disc}

\begin{figure}[]
	\figurenum{5}
	\fig{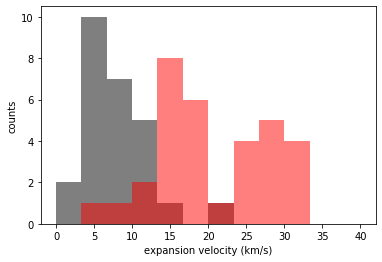}{.5\textwidth}{}
	\caption{ Histogram of the expansion velocity found in the previous studies of \cite{devicente2016} and \cite{jewell1991} (gray) and this survey (red). We find more high-velocity outflows in our sample.  \label{histfig}}
\end{figure}

We have presented a study of 43 GHz SiO $J=1-0, v=0$ emission in the BAaDE survey. Our sample size of 90 sources expands the sample examined in previous studies of SiO $J=1-0, v=0$ maser emission (\citealt{devicente2016} was previously the largest with a total of 28 sources). For the composite sources, our profiles are consistent with those of \cite{devicente2016}; we see the maser component centered on the stellar velocity, and we find composite profiles to be common. We find a wider variety of expansion velocities for our stars, as shown in Figure \ref{histfig}. We also include one source from \cite{jewell1991} that was not observed in \cite{devicente2016} or our sample, OH 2.6-0.4. The expansion velocities of the \cite{devicente2016} sources were estimated by eye based on those published profiles. We share two sources in common with this sample: ad3a-15253 (OH44.8-2.3), which we identify as a maser source, and ce3a-00110 (VX Sgr); our profile with a broad base and a maser peak looks similar to that in \cite{devicente2016}, and our rough estimate of the expansion velocity of that source based on the profile in \cite{devicente2016} is the same, about 24 km s$^{-1}$. Our value is consistent with the measurement by \cite{jewell1991} of 24.1 km s$^{-1}$. The overall lower expansion velocities of \cite{devicente2016} compared to our sources could be explained by the fact that those sources were optically selected and not necessarily in the same iiia region of MSX color space as our sources, and therefore not at the same mass-loss stage.

We are able to derive expansion velocities using a thermal emission model fit to the 32 profiles with thermal components. The four sources that we have in common with \cite{CO1998} have expansion velocities similar to those of CO, which tentatively suggests that the thermal SiO profiles are characterized by the region that has reached terminal velocity, as are CO profiles. There could be some secondary contribution from the dust formation zone, where the winds are being accelerated, but profiles measured with integration times longer than in our survey will be required to constrain such contributions. Future high-resolution VLBI mapping of the ground-state maser position in the circumstellar envelope would shed light on the question of what outflow regime this transition is probing, and would allow for comparison to pumping models, as in \cite{SR2004}, for example. Future studies could also examine the time-domain features of the $v=0$ masers. In \cite{devicente2016} the sources that were followed up showed variability in the maser component that was not related to the optical or infrared variability of the stellar source.

We find that our sources are brighter than the main locus of the BAaDE survey and present no clustering in space. These properties, supplemented with a kinematic argument and some \textit{Gaia} distances, suggest that our sources are members of the disk of the Milky Way. This analysis suggests that we are seeing ground-state emission in nearby sources because it is too faint to see at greater distances, implying that deeper observations of more distant SiO sources will likely result in more detections of ground-state emission.

\acknowledgments

The authors thank Y.M. Pihlstr{\"o}m and L.O. Sjouwerman for their helpful comments.

This material is based upon work supported by the National Science Foundation Graduate Research Fellowship Program under Grant No. DGE-1650604, and by the  National Science Foundation under Grant Number 1518271 to UCLA. Any opinions, findings, and conclusions or recommendations expressed in this material are those of the authors and do not necessarily reflect the views of the National Science Foundation.

V.D. thanks the LSSTC Data Science Fellowship Program, which is funded by LSSTC, NSF Cybertraining Grant \#1829740, the Brinson Foundation, and the Moore Foundation; her participation in the program has benefited this work.

M.C.S. is partially supported by the Heising-Simons Foundation under grant \#2018- 0911.

Support for this work was provided by the NSF through the Grote Reber Fellowship Program administered by Associated Universities, Inc./National Radio Astronomy Observatory.

The National Radio Astronomy Observatory is a facility of the National Science Foundation operated under cooperative agreement by Associated Universities, Inc.

This research made use of the SIMBAD database, operated at CDS,  Strasbourg,  France.   This  research  made  use  of data  products  from  the  Midcourse  Space  Experiment. This publication makes use  of  data  products  from  the  Two  Micron  All  Sky Survey,  which  is  a  joint  project  of  the  University  of Massachusetts  and  the  Infrared  Processing  and  Analysis  Center/California  Institute  of  Technology,  funded by the National Aeronautics and Space Administration and the National Science Foundation. 

This work has made use of data from the European Space Agency (ESA) mission
{\it Gaia} (\url{https://www.cosmos.esa.int/gaia}), processed by the {\it Gaia}
Data Processing and Analysis Consortium (DPAC,
\url{https://www.cosmos.esa.int/web/gaia/dpac/consortium}). Funding for the DPAC
has been provided by national institutions, in particular the institutions
participating in the {\it Gaia} Multilateral Agreement.

\facility{VLA}
\software{Astropy (\citealt{astropy1} \& \citealt{astropy2}), Matplotlib \citep{matplotlib}, NumPy \citep{numpy}, pandas \citep{pandas},  SciPy \citep{scipy}}

\bibliographystyle{aasjournal}
\bibliography{dikemaserpaper}

\appendix
\figsetstart
\figsetnum{A1}
\figsettitle{Maser Spectra}

\figsetgrpstart
\figsetgrpnum{A1.1}
\figsetgrptitle{Spectrum of ad3a-00230
}
\figsetplot{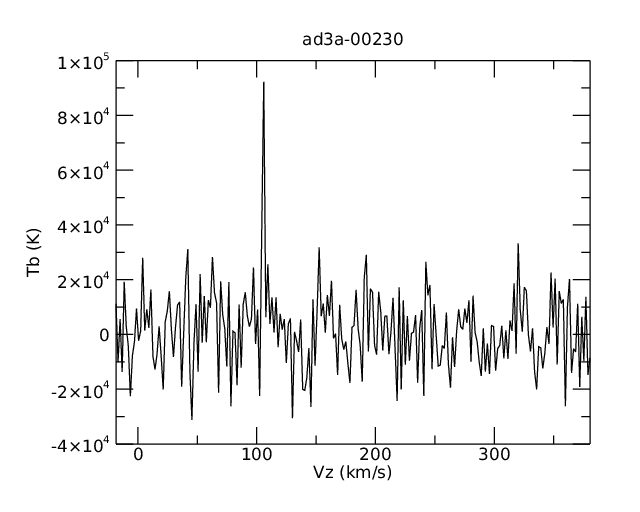}
\figsetgrpnote{Line profiles of the $v=0, J=1-0$ masers in our sample. $V_z$ is the local standard of rest velocity.}
\figsetgrpend

\figsetgrpstart
\figsetgrpnum{A1.2}
\figsetgrptitle{Spectrum of ad3a-00531
}
\figsetplot{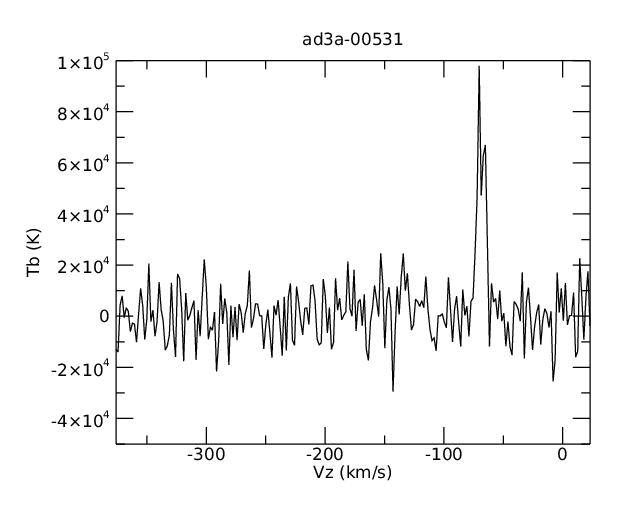}
\figsetgrpnote{Line profiles of the $v=0, J=1-0$ masers in our sample. $V_z$ is the local standard of rest velocity.}
\figsetgrpend

\figsetgrpstart
\figsetgrpnum{A1.3}
\figsetgrptitle{Spectrum of ad3a-00562
}
\figsetplot{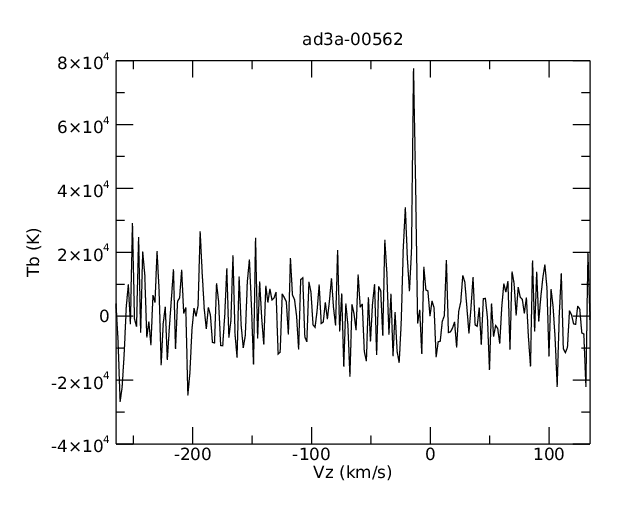}
\figsetgrpnote{Line profiles of the $v=0, J=1-0$ masers in our sample. $V_z$ is the local standard of rest velocity.}
\figsetgrpend

\figsetgrpstart
\figsetgrpnum{A1.4}
\figsetgrptitle{Spectrum of ad3a-00582
}
\figsetplot{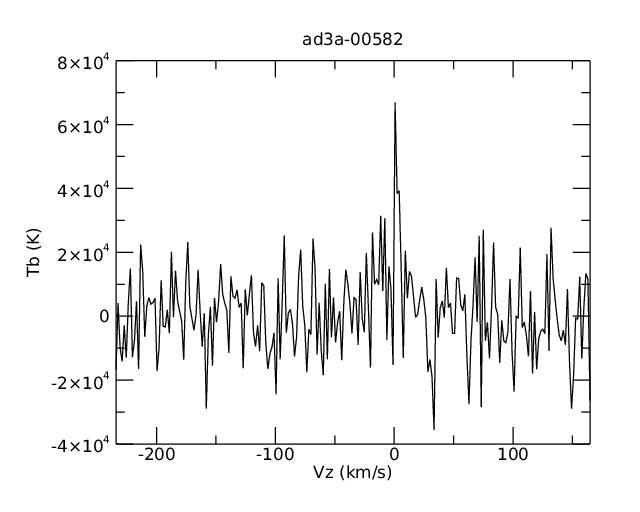}
\figsetgrpnote{Line profiles of the $v=0, J=1-0$ masers in our sample. $V_z$ is the local standard of rest velocity.}
\figsetgrpend

\figsetgrpstart
\figsetgrpnum{A1.5}
\figsetgrptitle{Spectrum of ad3a-01244
}
\figsetplot{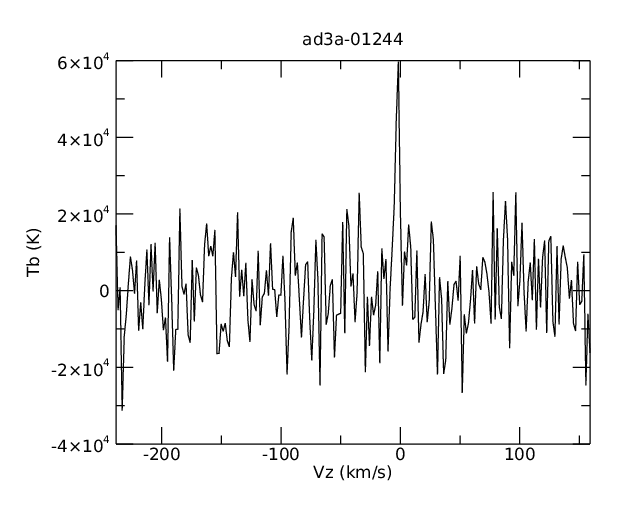}
\figsetgrpnote{Line profiles of the $v=0, J=1-0$ masers in our sample. $V_z$ is the local standard of rest velocity.}
\figsetgrpend

\figsetgrpstart
\figsetgrpnum{A1.6}
\figsetgrptitle{Spectrum of ad3a-01758
}
\figsetplot{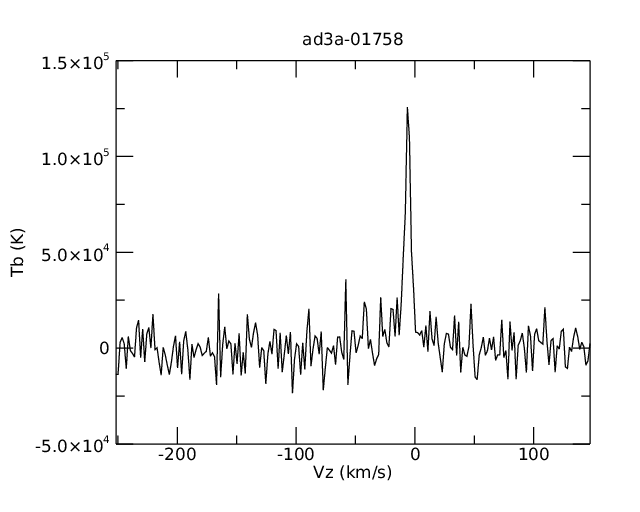}
\figsetgrpnote{Line profiles of the $v=0, J=1-0$ masers in our sample. $V_z$ is the local standard of rest velocity.}
\figsetgrpend

\figsetgrpstart
\figsetgrpnum{A1.7}
\figsetgrptitle{Spectrum of ad3a-01815
}
\figsetplot{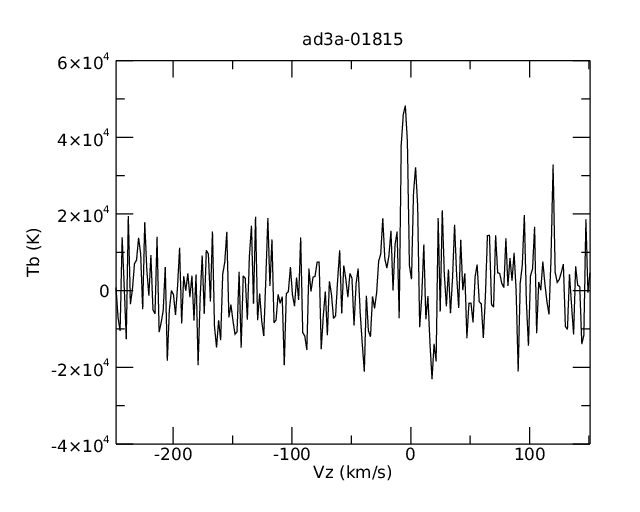}
\figsetgrpnote{Line profiles of the $v=0, J=1-0$ masers in our sample. $V_z$ is the local standard of rest velocity.}
\figsetgrpend

\figsetgrpstart
\figsetgrpnum{A1.8}
\figsetgrptitle{Spectrum of ad3a-01975
}
\figsetplot{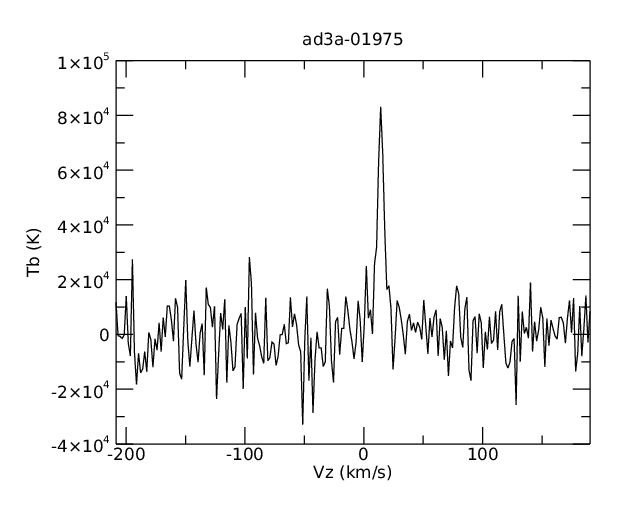}
\figsetgrpnote{Line profiles of the $v=0, J=1-0$ masers in our sample. $V_z$ is the local standard of rest velocity.}
\figsetgrpend

\figsetgrpstart
\figsetgrpnum{A1.9}
\figsetgrptitle{Spectrum of ad3a-02502
}
\figsetplot{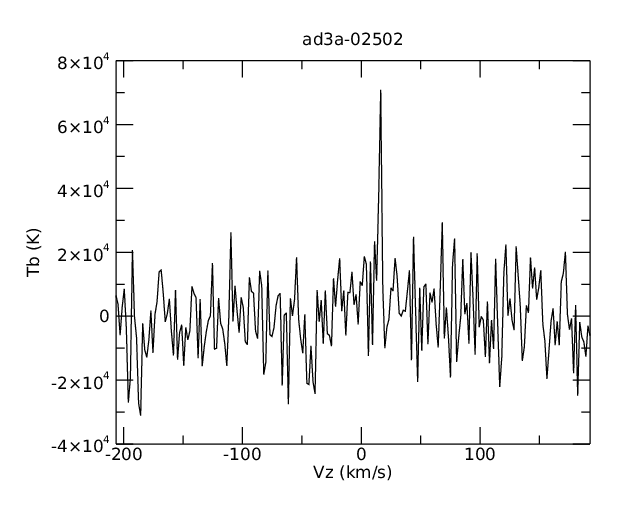}
\figsetgrpnote{Line profiles of the $v=0, J=1-0$ masers in our sample. $V_z$ is the local standard of rest velocity.}
\figsetgrpend

\figsetgrpstart
\figsetgrpnum{A1.10}
\figsetgrptitle{Spectrum of ad3a-04289
}
\figsetplot{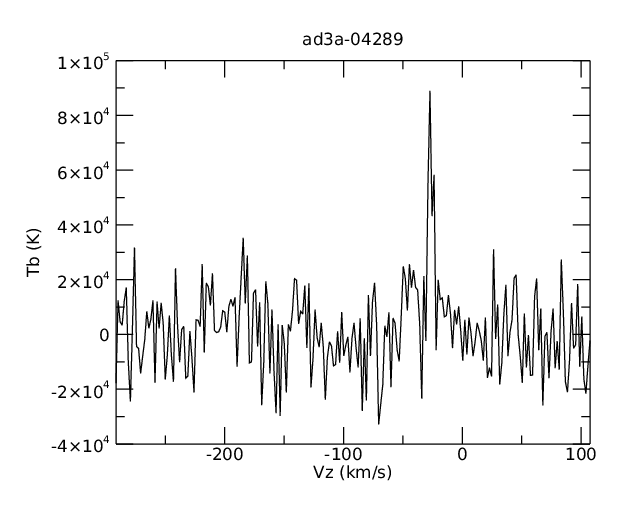}
\figsetgrpnote{Line profiles of the $v=0, J=1-0$ masers in our sample. $V_z$ is the local standard of rest velocity.}
\figsetgrpend

\figsetgrpstart
\figsetgrpnum{A1.11}
\figsetgrptitle{Spectrum of ad3a-04410
}
\figsetplot{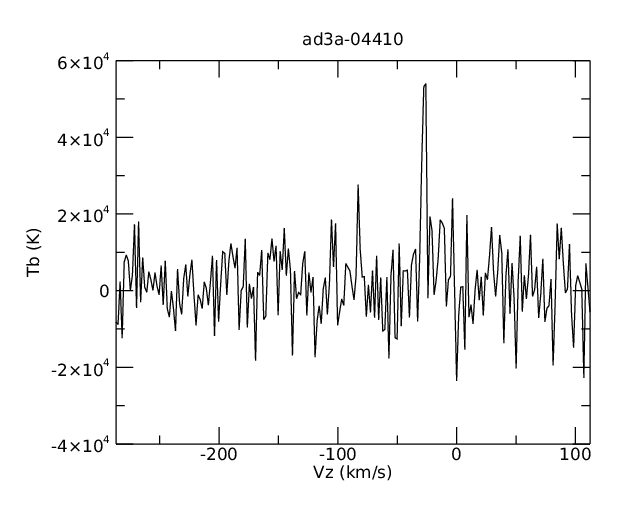}
\figsetgrpnote{Line profiles of the $v=0, J=1-0$ masers in our sample. $V_z$ is the local standard of rest velocity.}
\figsetgrpend

\figsetgrpstart
\figsetgrpnum{A1.12}
\figsetgrptitle{Spectrum of ad3a-04942
}
\figsetplot{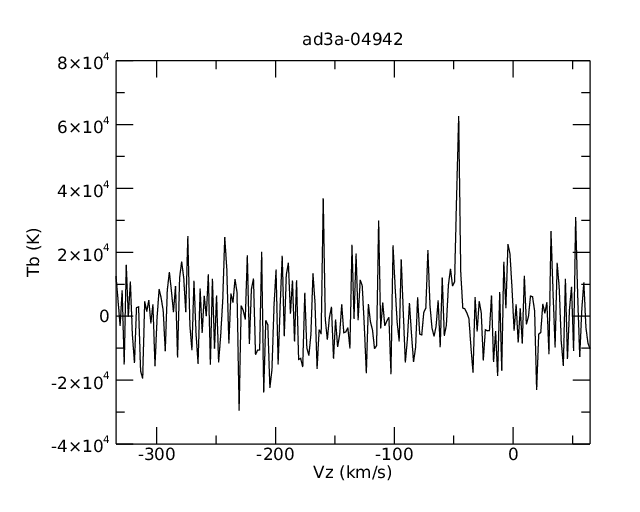}
\figsetgrpnote{Line profiles of the $v=0, J=1-0$ masers in our sample. $V_z$ is the local standard of rest velocity.}
\figsetgrpend

\figsetgrpstart
\figsetgrpnum{A1.13}
\figsetgrptitle{Spectrum of ad3a-04944
}
\figsetplot{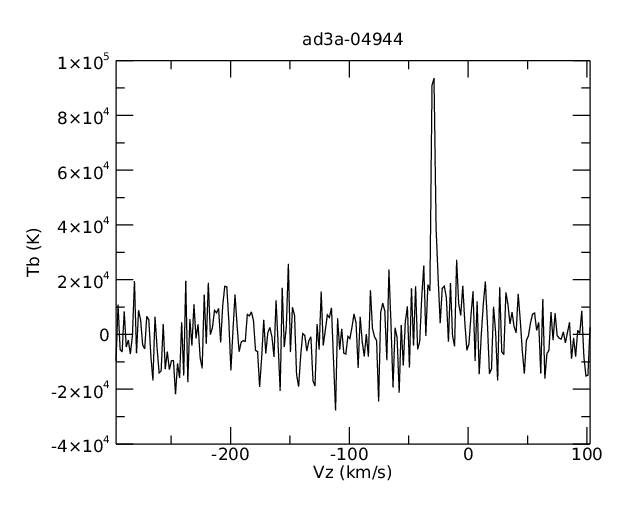}
\figsetgrpnote{Line profiles of the $v=0, J=1-0$ masers in our sample. $V_z$ is the local standard of rest velocity.}
\figsetgrpend

\figsetgrpstart
\figsetgrpnum{A1.14}
\figsetgrptitle{Spectrum of ad3a-05408
}
\figsetplot{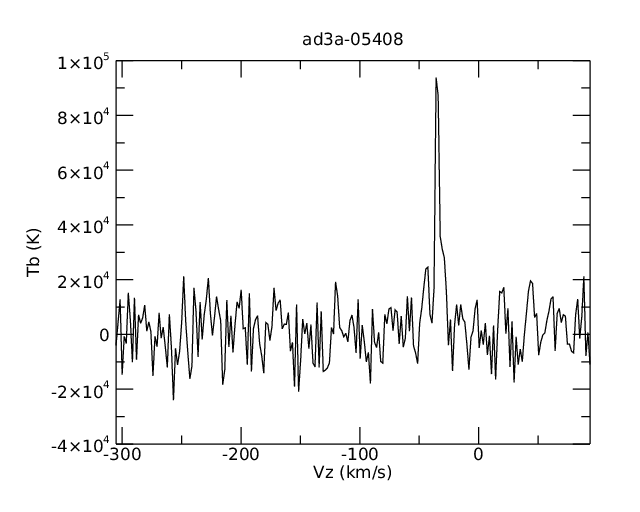}
\figsetgrpnote{Line profiles of the $v=0, J=1-0$ masers in our sample. $V_z$ is the local standard of rest velocity.}
\figsetgrpend

\figsetgrpstart
\figsetgrpnum{A1.15}
\figsetgrptitle{Spectrum of ad3a-05488
}
\figsetplot{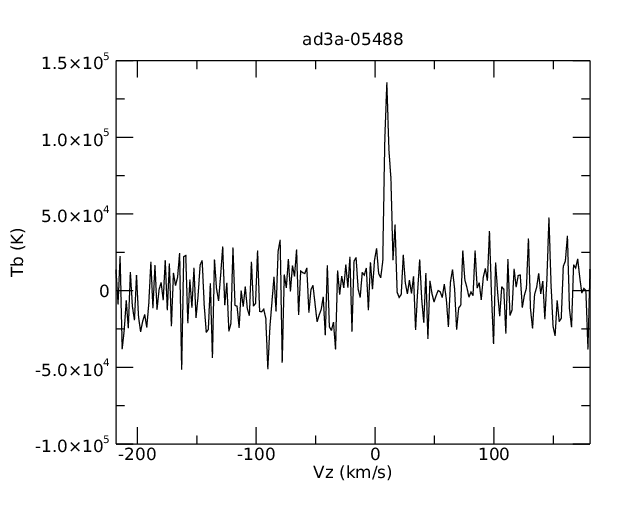}
\figsetgrpnote{Line profiles of the $v=0, J=1-0$ masers in our sample. $V_z$ is the local standard of rest velocity.}
\figsetgrpend

\figsetgrpstart
\figsetgrpnum{A1.16}
\figsetgrptitle{Spectrum of ad3a-05782
}
\figsetplot{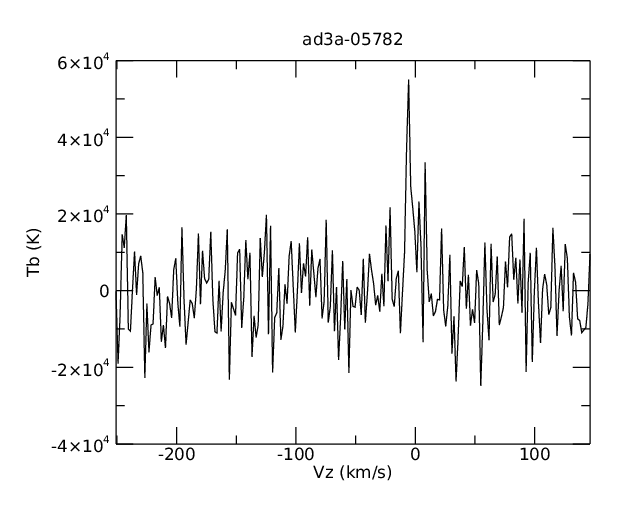}
\figsetgrpnote{Line profiles of the $v=0, J=1-0$ masers in our sample. $V_z$ is the local standard of rest velocity.}
\figsetgrpend

\figsetgrpstart
\figsetgrpnum{A1.17}
\figsetgrptitle{Spectrum of ad3a-06568
}
\figsetplot{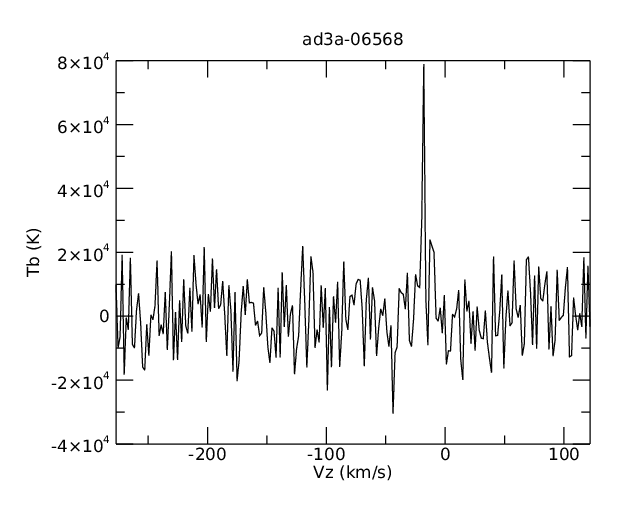}
\figsetgrpnote{Line profiles of the $v=0, J=1-0$ masers in our sample. $V_z$ is the local standard of rest velocity.}
\figsetgrpend

\figsetgrpstart
\figsetgrpnum{A1.18}
\figsetgrptitle{Spectrum of ad3a-07151
}
\figsetplot{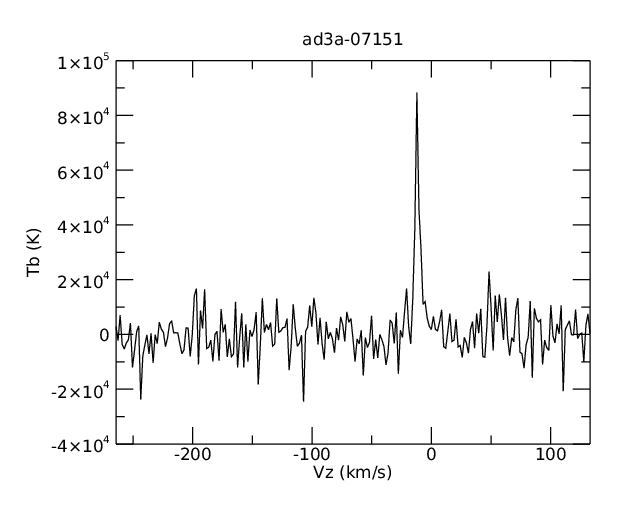}
\figsetgrpnote{Line profiles of the $v=0, J=1-0$ masers in our sample. $V_z$ is the local standard of rest velocity.}
\figsetgrpend

\figsetgrpstart
\figsetgrpnum{A1.19}
\figsetgrptitle{Spectrum of ad3a-08004
}
\figsetplot{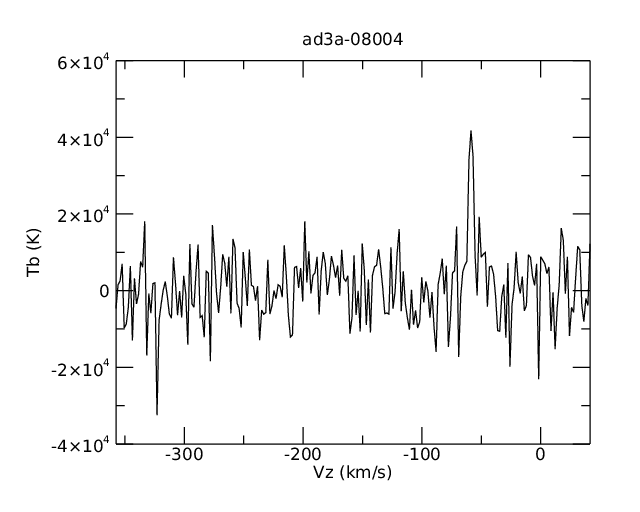}
\figsetgrpnote{Line profiles of the $v=0, J=1-0$ masers in our sample. $V_z$ is the local standard of rest velocity.}
\figsetgrpend

\figsetgrpstart
\figsetgrpnum{A1.20}
\figsetgrptitle{Spectrum of ad3a-08141
}
\figsetplot{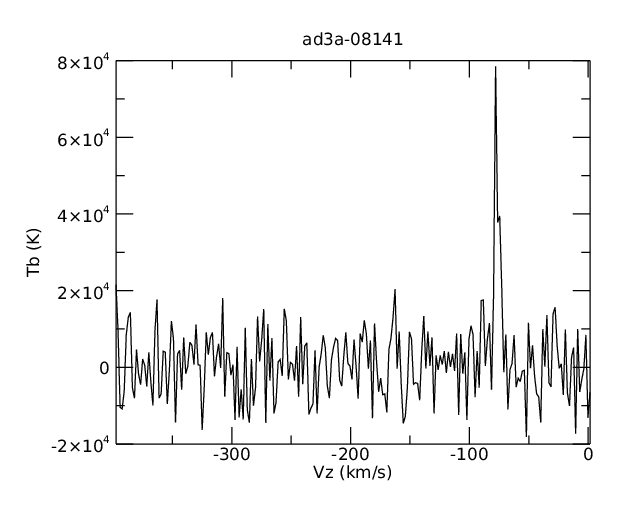}
\figsetgrpnote{Line profiles of the $v=0, J=1-0$ masers in our sample. $V_z$ is the local standard of rest velocity.}
\figsetgrpend

\figsetgrpstart
\figsetgrpnum{A1.21}
\figsetgrptitle{Spectrum of ad3a-08264
}
\figsetplot{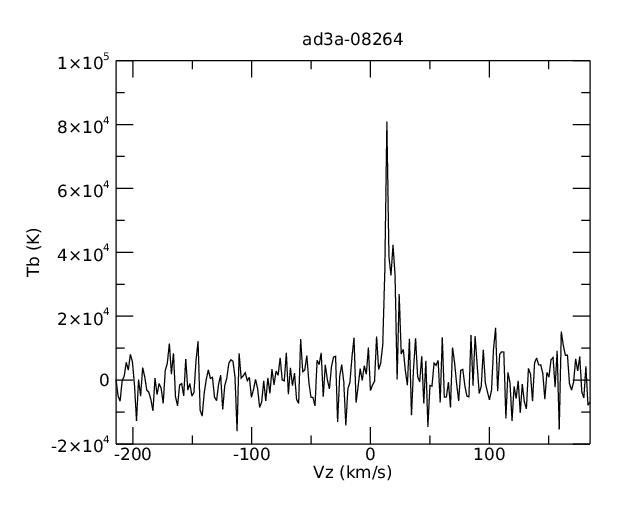}
\figsetgrpnote{Line profiles of the $v=0, J=1-0$ masers in our sample. $V_z$ is the local standard of rest velocity.}
\figsetgrpend

\figsetgrpstart
\figsetgrpnum{A1.22}
\figsetgrptitle{Spectrum of ad3a-08428
}
\figsetplot{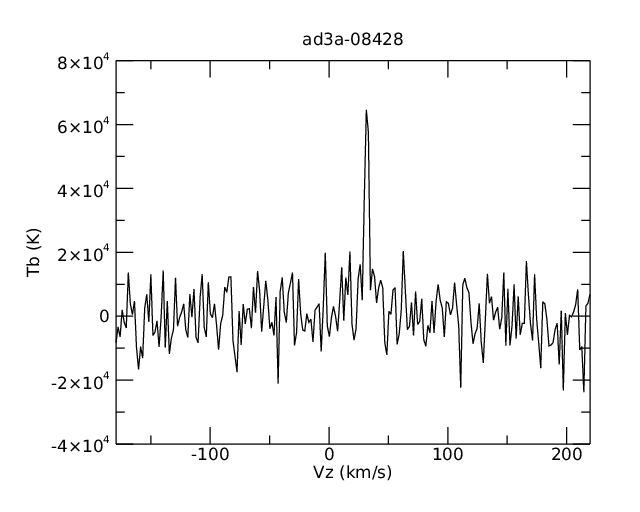}
\figsetgrpnote{Line profiles of the $v=0, J=1-0$ masers in our sample. $V_z$ is the local standard of rest velocity.}
\figsetgrpend

\figsetgrpstart
\figsetgrpnum{A1.23}
\figsetgrptitle{Spectrum of ad3a-08663
}
\figsetplot{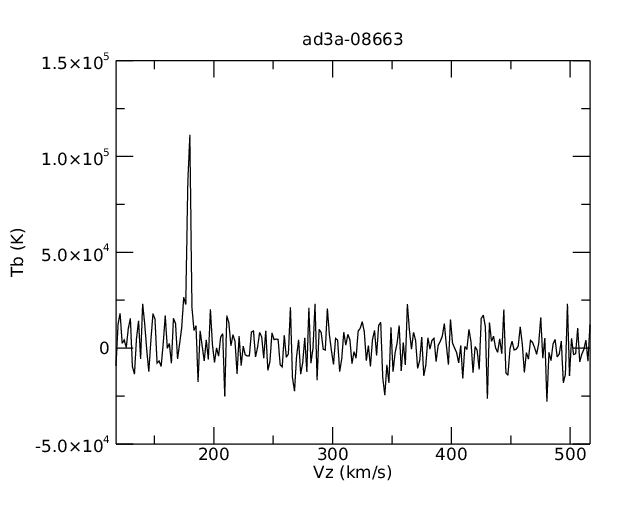}
\figsetgrpnote{Line profiles of the $v=0, J=1-0$ masers in our sample. $V_z$ is the local standard of rest velocity.}
\figsetgrpend

\figsetgrpstart
\figsetgrpnum{A1.24}
\figsetgrptitle{Spectrum of ad3a-09257
}
\figsetplot{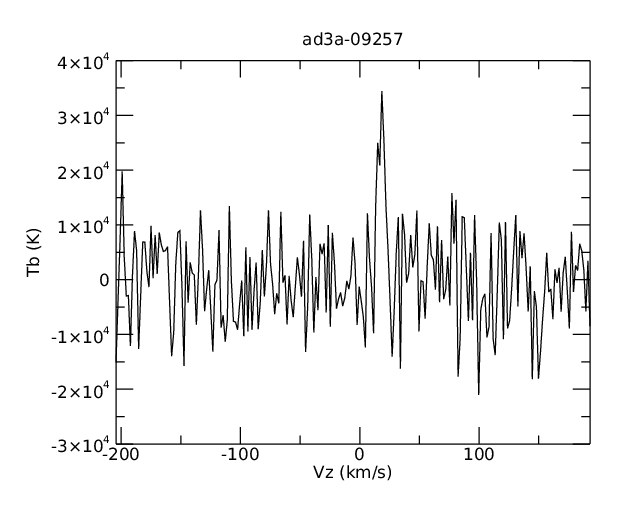}
\figsetgrpnote{Line profiles of the $v=0, J=1-0$ masers in our sample. $V_z$ is the local standard of rest velocity.}
\figsetgrpend

\figsetgrpstart
\figsetgrpnum{A1.25}
\figsetgrptitle{Spectrum of ad3a-10192
}
\figsetplot{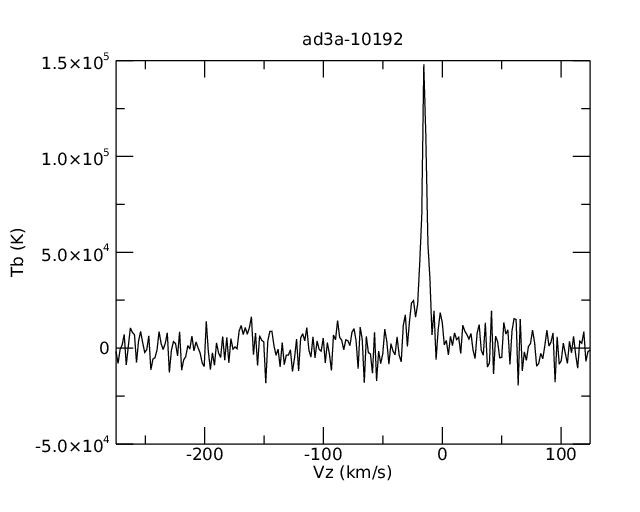}
\figsetgrpnote{Line profiles of the $v=0, J=1-0$ masers in our sample. $V_z$ is the local standard of rest velocity.}
\figsetgrpend

\figsetgrpstart
\figsetgrpnum{A1.26}
\figsetgrptitle{Spectrum of ad3a-10275
}
\figsetplot{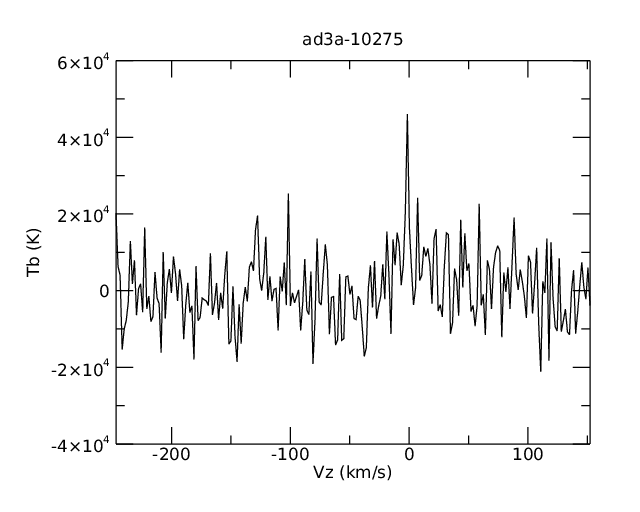}
\figsetgrpnote{Line profiles of the $v=0, J=1-0$ masers in our sample. $V_z$ is the local standard of rest velocity.}
\figsetgrpend

\figsetgrpstart
\figsetgrpnum{A1.27}
\figsetgrptitle{Spectrum of ad3a-10590
}
\figsetplot{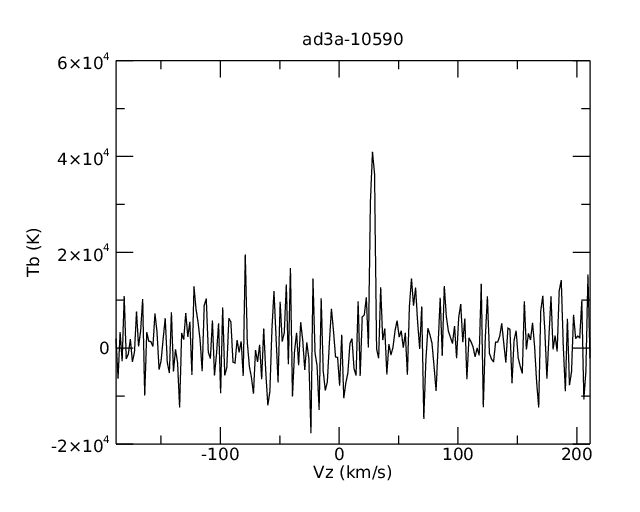}
\figsetgrpnote{Line profiles of the $v=0, J=1-0$ masers in our sample. $V_z$ is the local standard of rest velocity.}
\figsetgrpend

\figsetgrpstart
\figsetgrpnum{A1.28}
\figsetgrptitle{Spectrum of ad3a-11089
}
\figsetplot{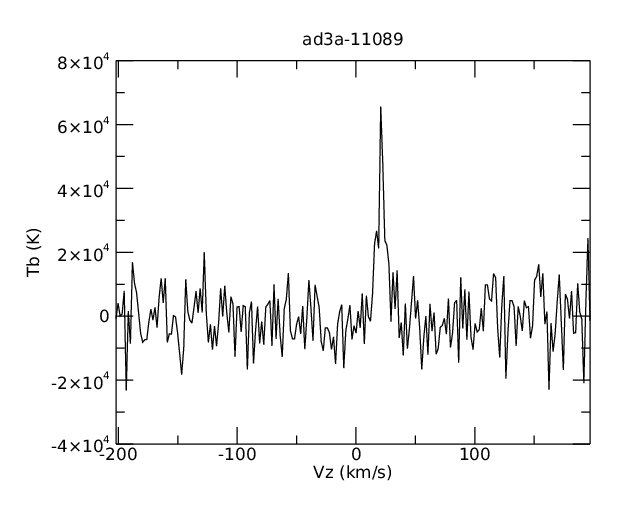}
\figsetgrpnote{Line profiles of the $v=0, J=1-0$ masers in our sample. $V_z$ is the local standard of rest velocity.}
\figsetgrpend

\figsetgrpstart
\figsetgrpnum{A1.29}
\figsetgrptitle{Spectrum of ad3a-11091
}
\figsetplot{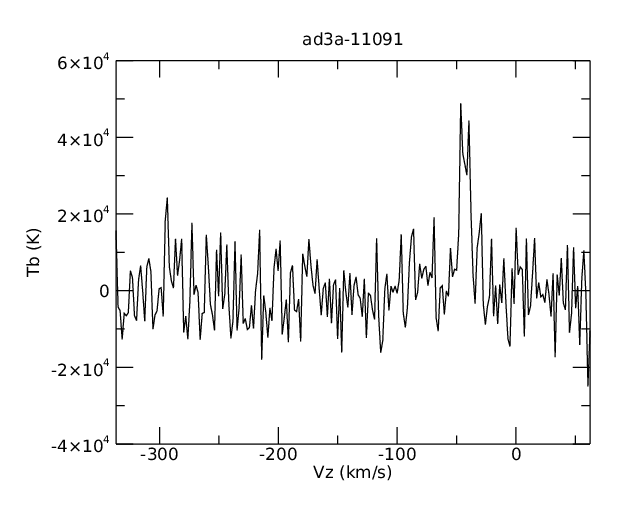}
\figsetgrpnote{Line profiles of the $v=0, J=1-0$ masers in our sample. $V_z$ is the local standard of rest velocity.}
\figsetgrpend

\figsetgrpstart
\figsetgrpnum{A1.30}
\figsetgrptitle{Spectrum of ad3a-11358
}
\figsetplot{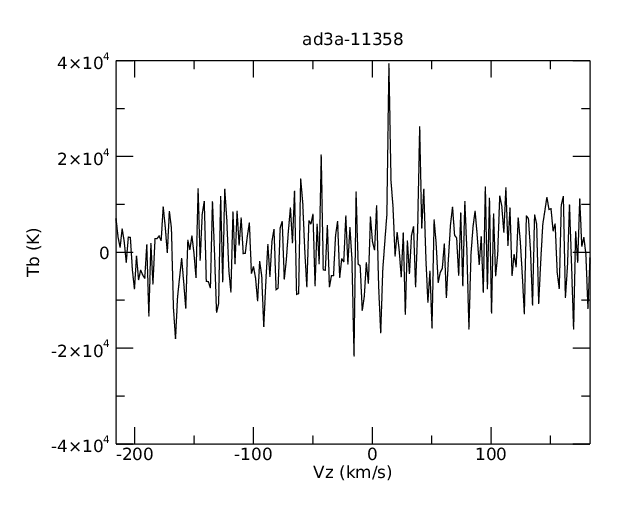}
\figsetgrpnote{Line profiles of the $v=0, J=1-0$ masers in our sample. $V_z$ is the local standard of rest velocity.}
\figsetgrpend

\figsetgrpstart
\figsetgrpnum{A1.31}
\figsetgrptitle{Spectrum of ad3a-11369
}
\figsetplot{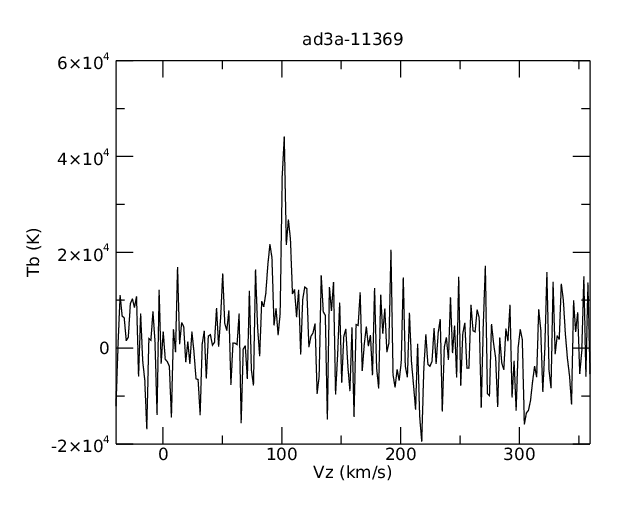}
\figsetgrpnote{Line profiles of the $v=0, J=1-0$ masers in our sample. $V_z$ is the local standard of rest velocity.}
\figsetgrpend

\figsetgrpstart
\figsetgrpnum{A1.32}
\figsetgrptitle{Spectrum of ad3a-11532
}
\figsetplot{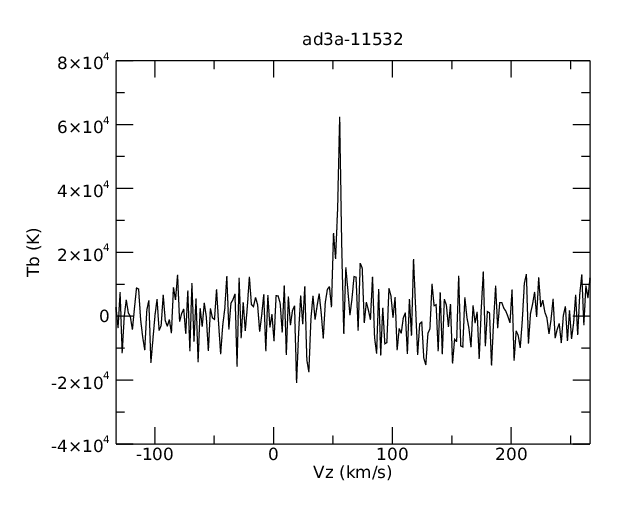}
\figsetgrpnote{Line profiles of the $v=0, J=1-0$ masers in our sample. $V_z$ is the local standard of rest velocity.}
\figsetgrpend

\figsetgrpstart
\figsetgrpnum{A1.33}
\figsetgrptitle{Spectrum of ad3a-12084
}
\figsetplot{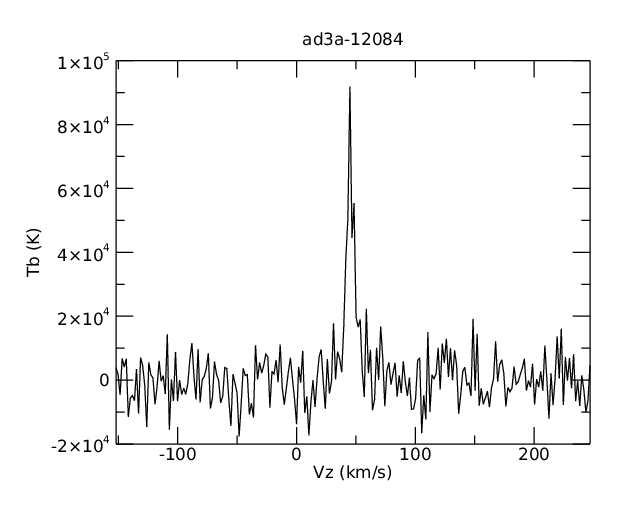}
\figsetgrpnote{Line profiles of the $v=0, J=1-0$ masers in our sample. $V_z$ is the local standard of rest velocity.}
\figsetgrpend

\figsetgrpstart
\figsetgrpnum{A1.34}
\figsetgrptitle{Spectrum of ad3a-12356
}
\figsetplot{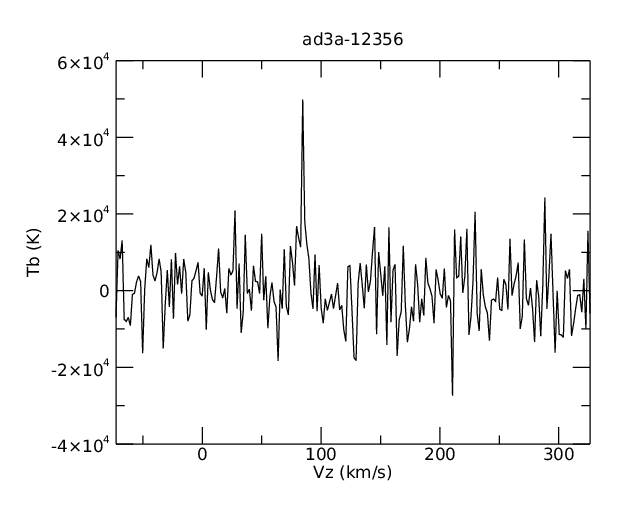}
\figsetgrpnote{Line profiles of the $v=0, J=1-0$ masers in our sample. $V_z$ is the local standard of rest velocity.}
\figsetgrpend

\figsetgrpstart
\figsetgrpnum{A1.35}
\figsetgrptitle{Spectrum of ad3a-12687
}
\figsetplot{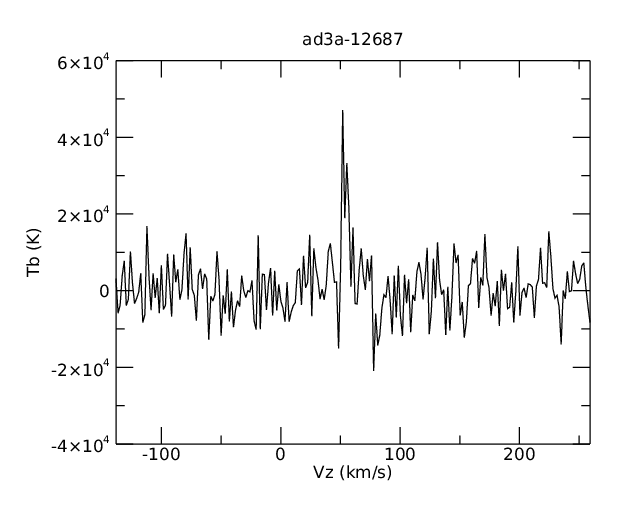}
\figsetgrpnote{Line profiles of the $v=0, J=1-0$ masers in our sample. $V_z$ is the local standard of rest velocity.}
\figsetgrpend

\figsetgrpstart
\figsetgrpnum{A1.36}
\figsetgrptitle{Spectrum of ad3a-12759
}
\figsetplot{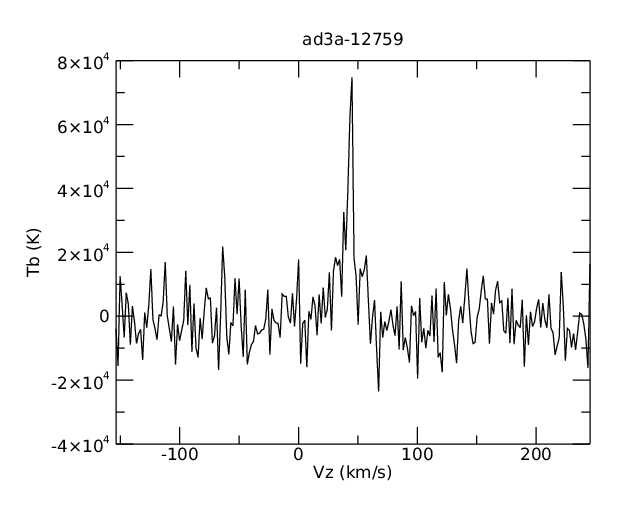}
\figsetgrpnote{Line profiles of the $v=0, J=1-0$ masers in our sample. $V_z$ is the local standard of rest velocity.}
\figsetgrpend

\figsetgrpstart
\figsetgrpnum{A1.37}
\figsetgrptitle{Spectrum of ad3a-13063
}
\figsetplot{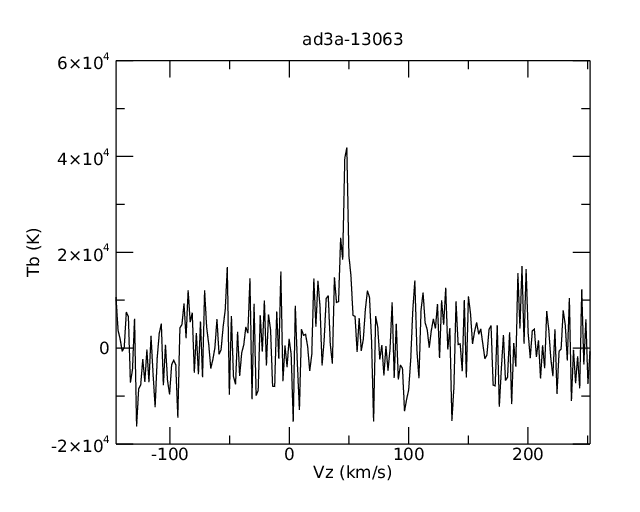}
\figsetgrpnote{Line profiles of the $v=0, J=1-0$ masers in our sample. $V_z$ is the local standard of rest velocity.}
\figsetgrpend

\figsetgrpstart
\figsetgrpnum{A1.38}
\figsetgrptitle{Spectrum of ad3a-13192
}
\figsetplot{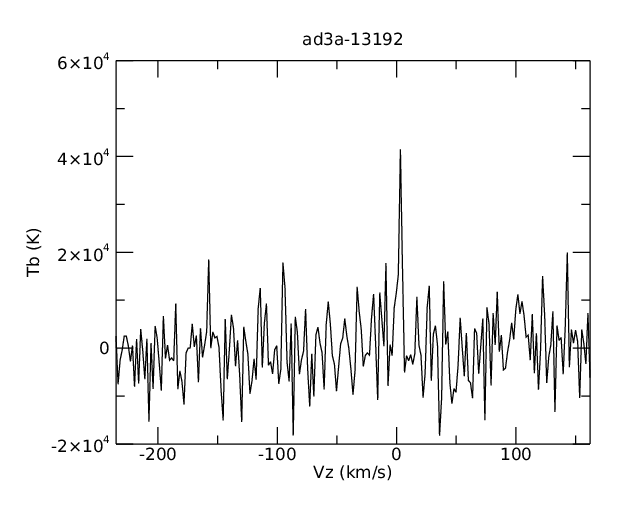}
\figsetgrpnote{Line profiles of the $v=0, J=1-0$ masers in our sample. $V_z$ is the local standard of rest velocity.}
\figsetgrpend

\figsetgrpstart
\figsetgrpnum{A1.39}
\figsetgrptitle{Spectrum of ad3a-13249
}
\figsetplot{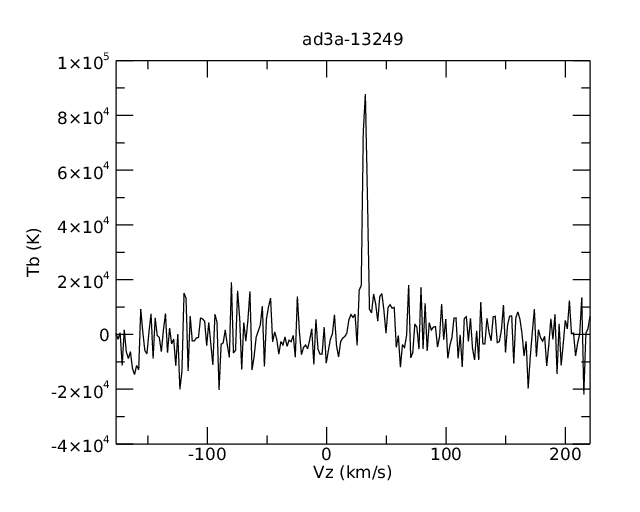}
\figsetgrpnote{Line profiles of the $v=0, J=1-0$ masers in our sample. $V_z$ is the local standard of rest velocity.}
\figsetgrpend

\figsetgrpstart
\figsetgrpnum{A1.40}
\figsetgrptitle{Spectrum of ad3a-13536
}
\figsetplot{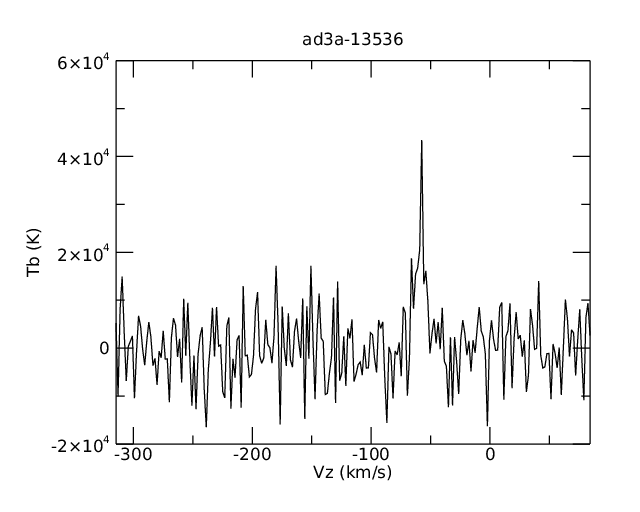}
\figsetgrpnote{Line profiles of the $v=0, J=1-0$ masers in our sample. $V_z$ is the local standard of rest velocity.}
\figsetgrpend

\figsetgrpstart
\figsetgrpnum{A1.41}
\figsetgrptitle{Spectrum of ad3a-13641
}
\figsetplot{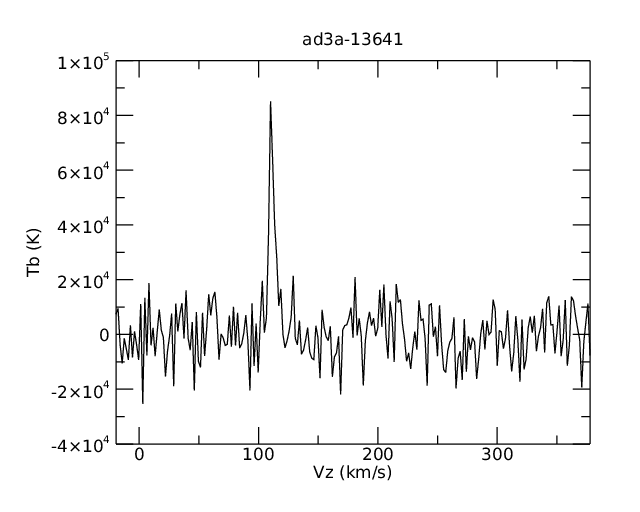}
\figsetgrpnote{Line profiles of the $v=0, J=1-0$ masers in our sample. $V_z$ is the local standard of rest velocity.}
\figsetgrpend

\figsetgrpstart
\figsetgrpnum{A1.42}
\figsetgrptitle{Spectrum of ad3a-14145
}
\figsetplot{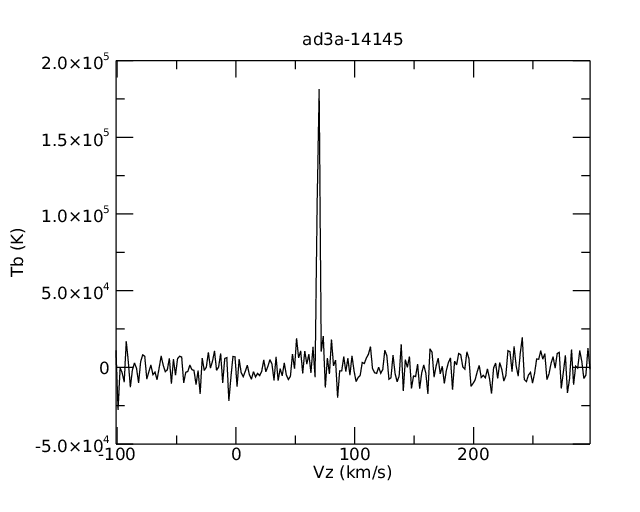}
\figsetgrpnote{Line profiles of the $v=0, J=1-0$ masers in our sample. $V_z$ is the local standard of rest velocity.}
\figsetgrpend

\figsetgrpstart
\figsetgrpnum{A1.43}
\figsetgrptitle{Spectrum of ad3a-15065
}
\figsetplot{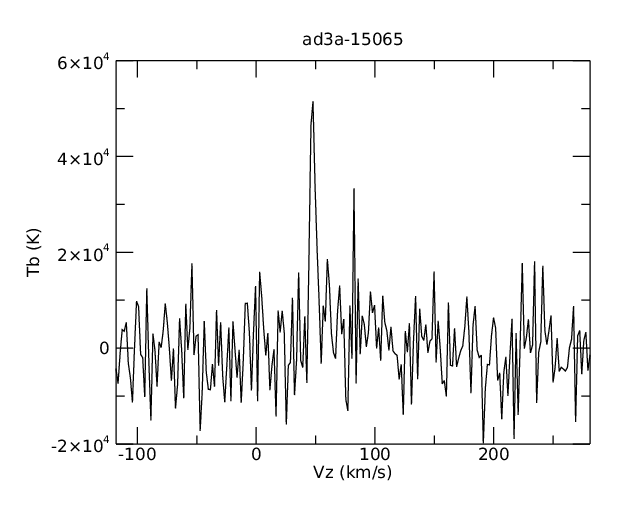}
\figsetgrpnote{Line profiles of the $v=0, J=1-0$ masers in our sample. $V_z$ is the local standard of rest velocity.}
\figsetgrpend

\figsetgrpstart
\figsetgrpnum{A1.44}
\figsetgrptitle{Spectrum of ad3a-15253
}
\figsetplot{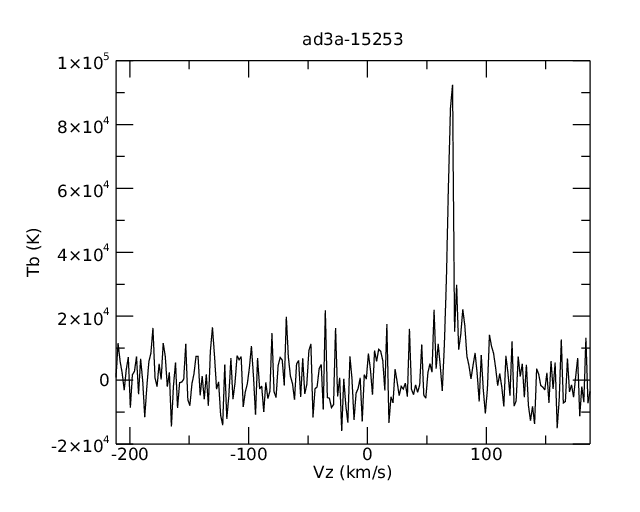}
\figsetgrpnote{Line profiles of the $v=0, J=1-0$ masers in our sample. $V_z$ is the local standard of rest velocity.}
\figsetgrpend

\figsetgrpstart
\figsetgrpnum{A1.45}
\figsetgrptitle{Spectrum of ad3a-15268
}
\figsetplot{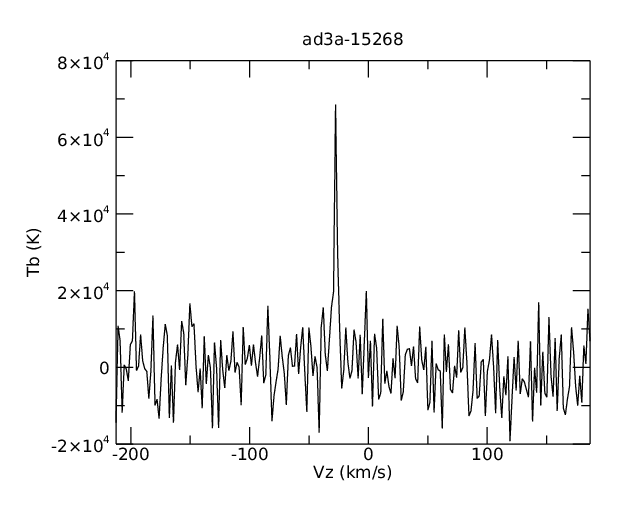}
\figsetgrpnote{Line profiles of the $v=0, J=1-0$ masers in our sample. $V_z$ is the local standard of rest velocity.}
\figsetgrpend

\figsetgrpstart
\figsetgrpnum{A1.46}
\figsetgrptitle{Spectrum of ad3a-17013
}
\figsetplot{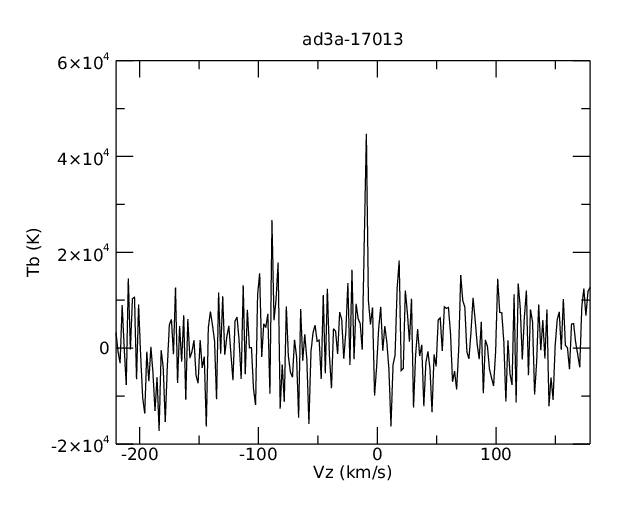}
\figsetgrpnote{Line profiles of the $v=0, J=1-0$ masers in our sample. $V_z$ is the local standard of rest velocity.}
\figsetgrpend

\figsetgrpstart
\figsetgrpnum{A1.47}
\figsetgrptitle{Spectrum of ad3a-17084
}
\figsetplot{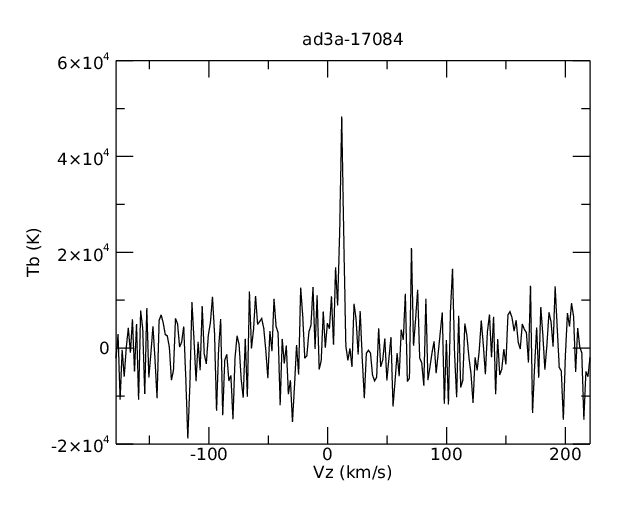}
\figsetgrpnote{Line profiles of the $v=0, J=1-0$ masers in our sample. $V_z$ is the local standard of rest velocity.}
\figsetgrpend

\figsetgrpstart
\figsetgrpnum{A1.48}
\figsetgrptitle{Spectrum of ad3a-17227
}
\figsetplot{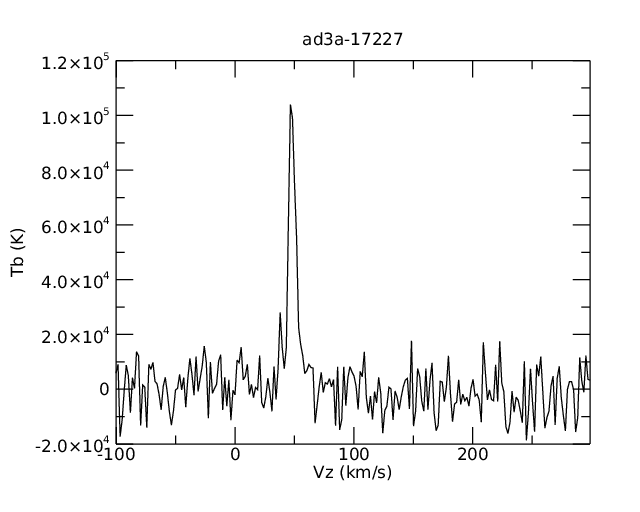}
\figsetgrpnote{Line profiles of the $v=0, J=1-0$ masers in our sample. $V_z$ is the local standard of rest velocity.}
\figsetgrpend

\figsetgrpstart
\figsetgrpnum{A1.49}
\figsetgrptitle{Spectrum of ae3a-00115
}
\figsetplot{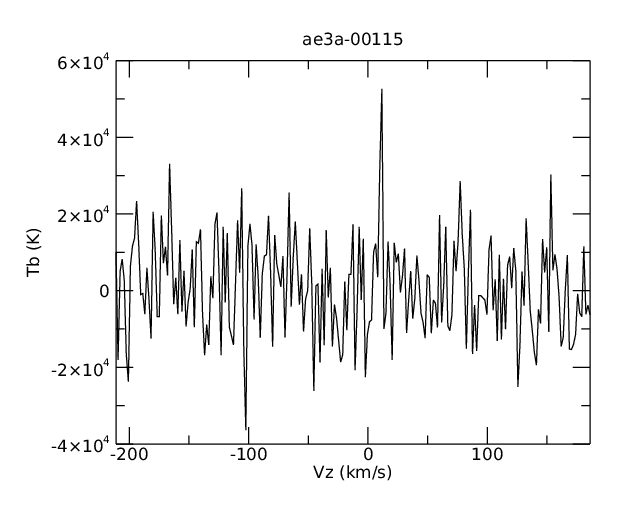}
\figsetgrpnote{Line profiles of the $v=0, J=1-0$ masers in our sample. $V_z$ is the local standard of rest velocity.}
\figsetgrpend

\figsetgrpstart
\figsetgrpnum{A1.50}
\figsetgrptitle{Spectrum of ae3a-00116
}
\figsetplot{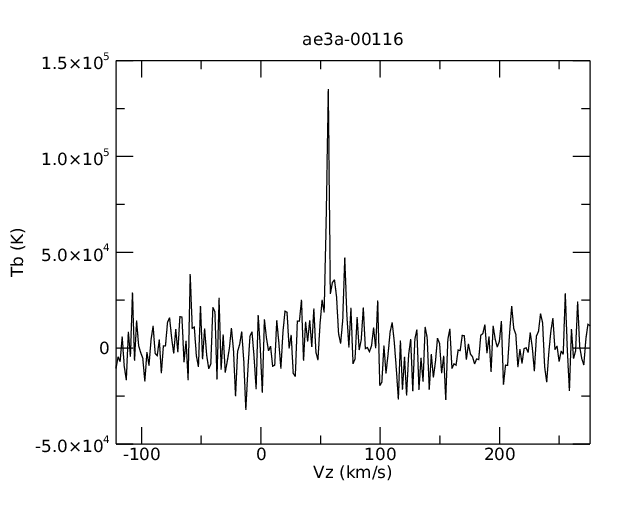}
\figsetgrpnote{Line profiles of the $v=0, J=1-0$ masers in our sample. $V_z$ is the local standard of rest velocity.}
\figsetgrpend

\figsetgrpstart
\figsetgrpnum{A1.51}
\figsetgrptitle{Spectrum of ae3a-00249
}
\figsetplot{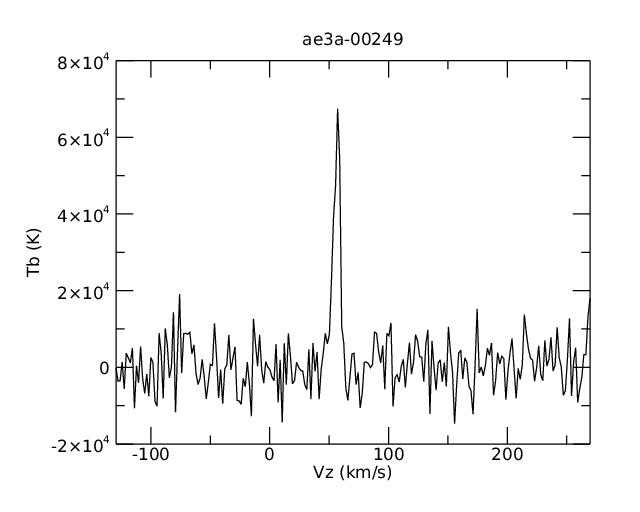}
\figsetgrpnote{Line profiles of the $v=0, J=1-0$ masers in our sample. $V_z$ is the local standard of rest velocity.}
\figsetgrpend

\figsetgrpstart
\figsetgrpnum{A1.52}
\figsetgrptitle{Spectrum of ae3a-00272
}
\figsetplot{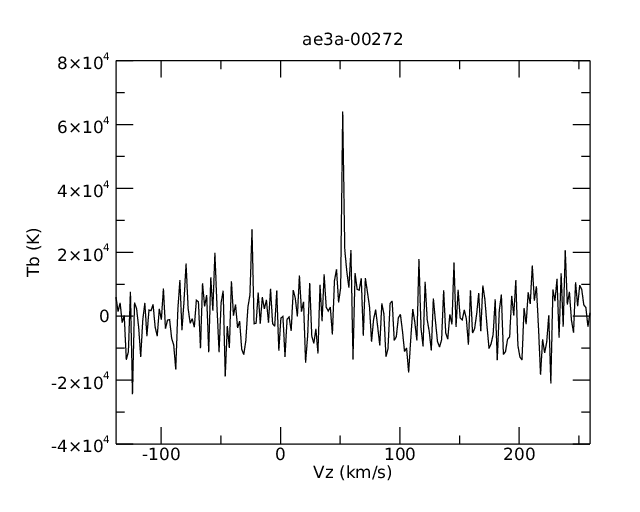}
\figsetgrpnote{Line profiles of the $v=0, J=1-0$ masers in our sample. $V_z$ is the local standard of rest velocity.}
\figsetgrpend

\figsetgrpstart
\figsetgrpnum{A1.53}
\figsetgrptitle{Spectrum of ae3a-03912
}
\figsetplot{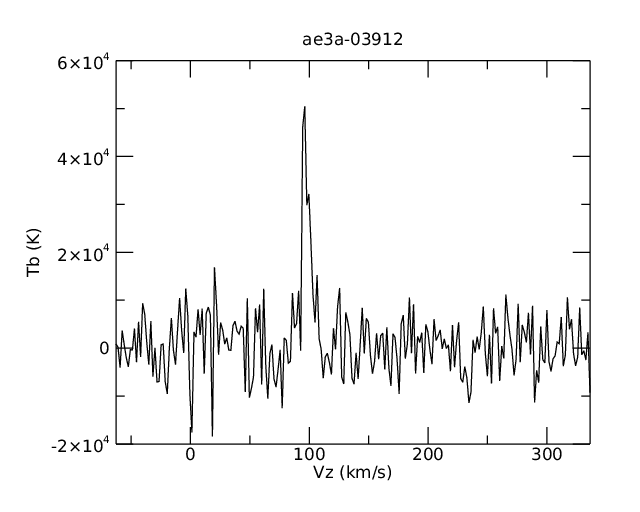}
\figsetgrpnote{Line profiles of the $v=0, J=1-0$ masers in our sample. $V_z$ is the local standard of rest velocity.}
\figsetgrpend

\figsetgrpstart
\figsetgrpnum{A1.54}
\figsetgrptitle{Spectrum of ce3a-00005
}
\figsetplot{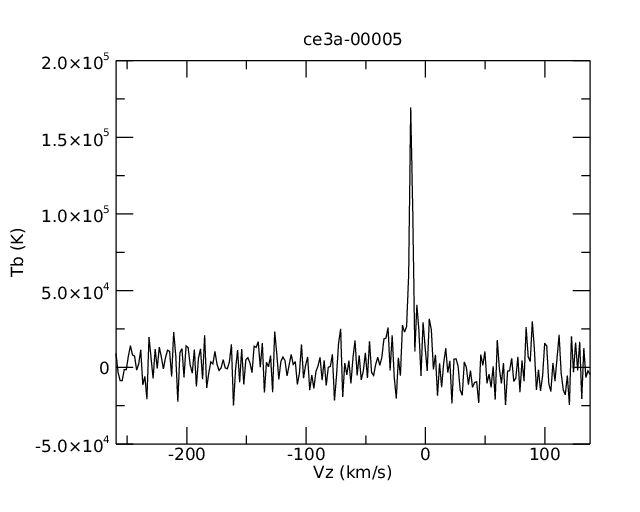}
\figsetgrpnote{Line profiles of the $v=0, J=1-0$ masers in our sample. $V_z$ is the local standard of rest velocity.}
\figsetgrpend

\figsetgrpstart
\figsetgrpnum{A1.55}
\figsetgrptitle{Spectrum of ce3a-00142
}
\figsetplot{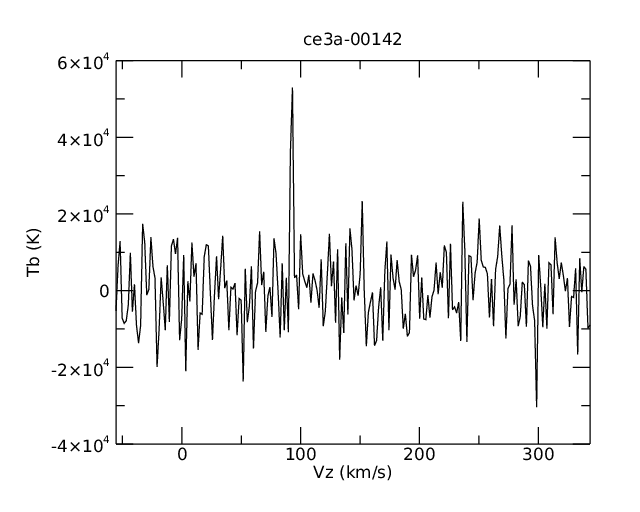}
\figsetgrpnote{Line profiles of the $v=0, J=1-0$ masers in our sample. $V_z$ is the local standard of rest velocity.}
\figsetgrpend

\figsetgrpstart
\figsetgrpnum{A1.56}
\figsetgrptitle{Spectrum of ce3a-00147
}
\figsetplot{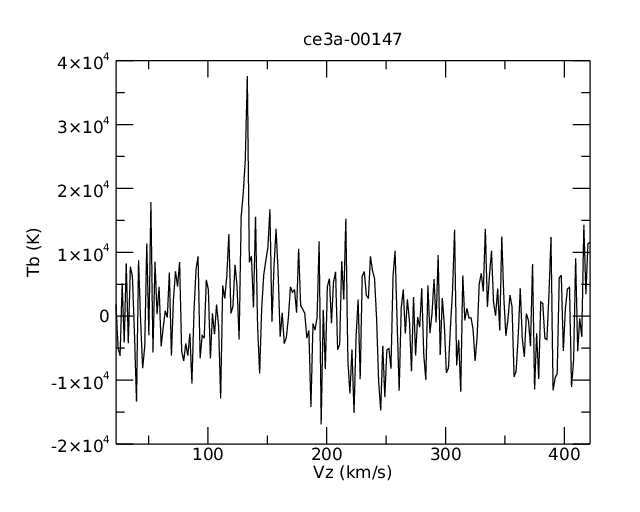}
\figsetgrpnote{Line profiles of the $v=0, J=1-0$ masers in our sample. $V_z$ is the local standard of rest velocity.}
\figsetgrpend

\figsetgrpstart
\figsetgrpnum{A1.57}
\figsetgrptitle{Spectrum of ce3a-00163
}
\figsetplot{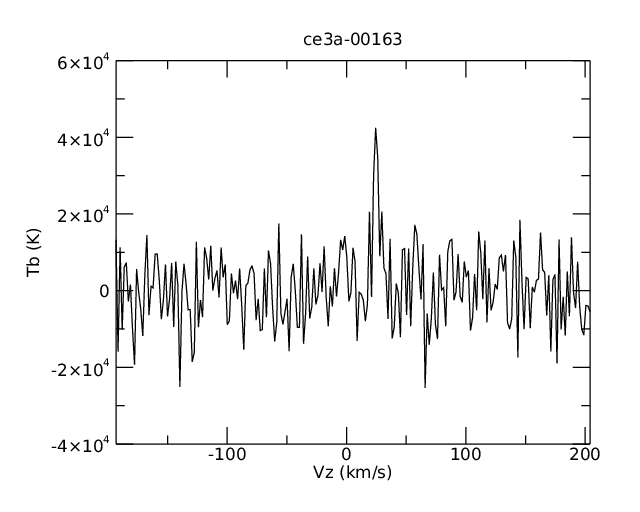}
\figsetgrpnote{Line profiles of the $v=0, J=1-0$ masers in our sample. $V_z$ is the local standard of rest velocity.}
\figsetgrpend

\figsetgrpstart
\figsetgrpnum{A1.58}
\figsetgrptitle{Spectrum of ce3a-00181
}
\figsetplot{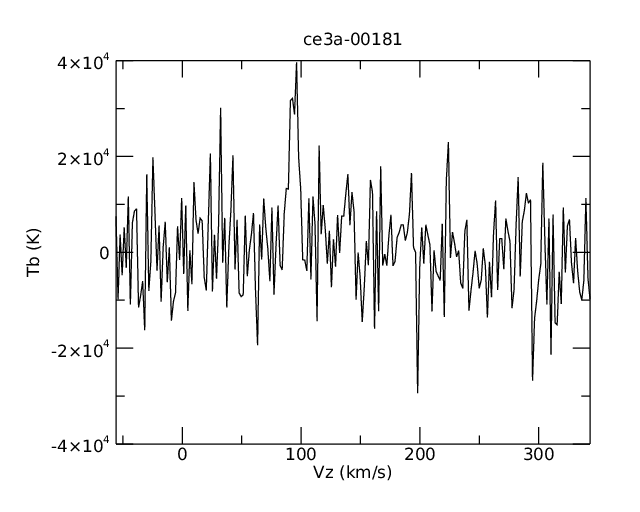}
\figsetgrpnote{Line profiles of the $v=0, J=1-0$ masers in our sample. $V_z$ is the local standard of rest velocity.}
\figsetgrpend

\figsetend

\begin{figure}[b]
\figurenum{A1}
\plotone{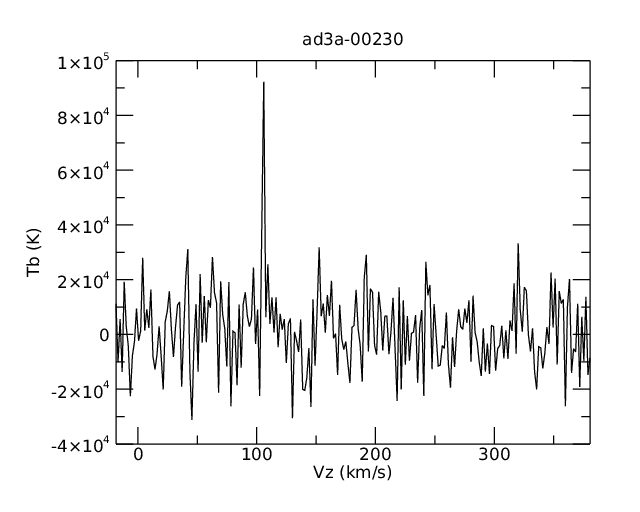}
\caption{Example profile of the $v=0, J=1-0$ maser line in source ad3a-00230. $V_z$ is the local standard of rest velocity. The complete figure set (58 maser-only spectra) is available in the online journal. \label{mfig}}
\end{figure}

\begin{figure}
	\figurenum{A2}
	\gridline{\fig{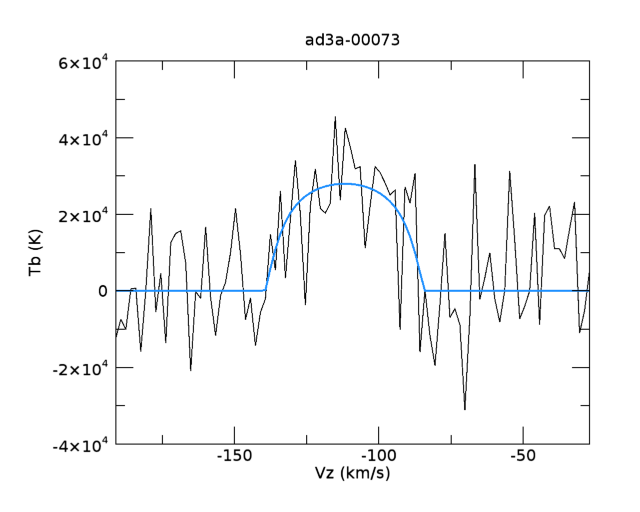}{.33\textwidth}{}
		\fig{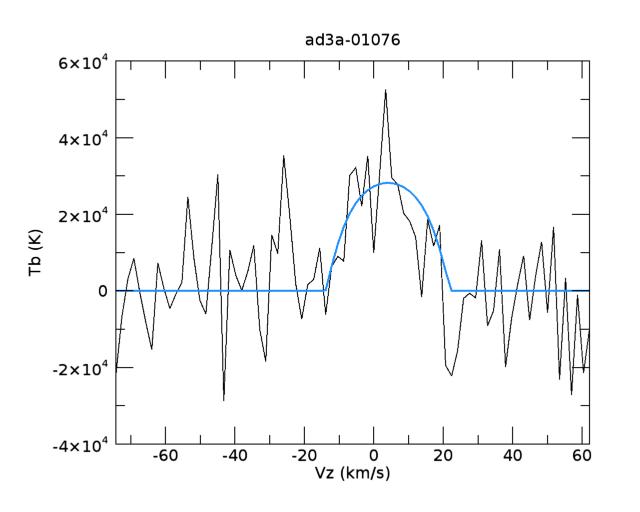}{.33\textwidth}{}
	\fig{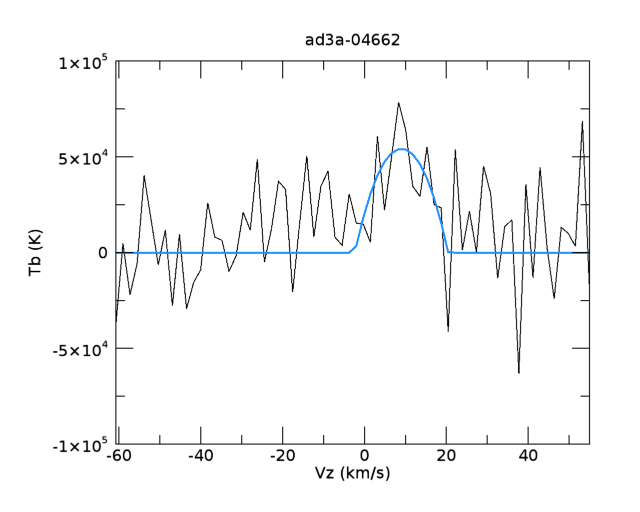}{.33\textwidth}{}}
		\gridline{\fig{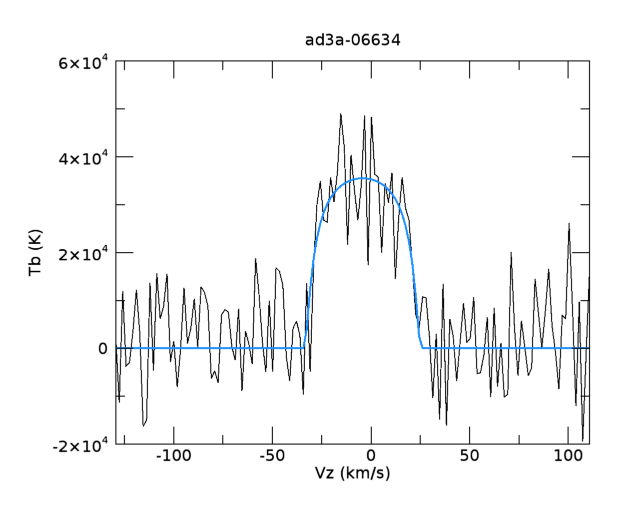}{.33\textwidth}{}
	\fig{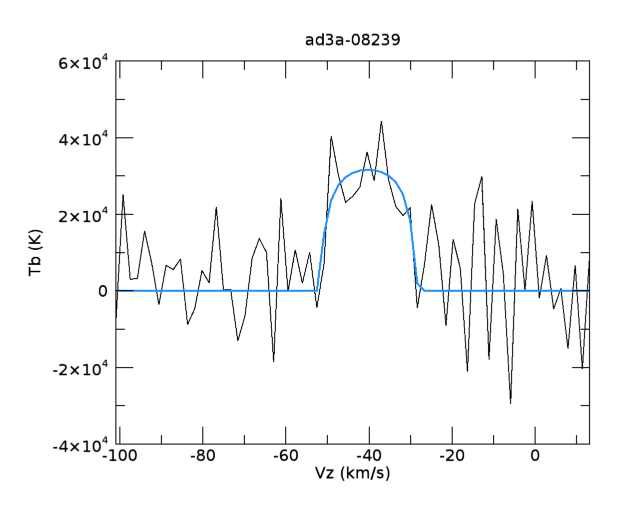}{.33\textwidth}{}
		\fig{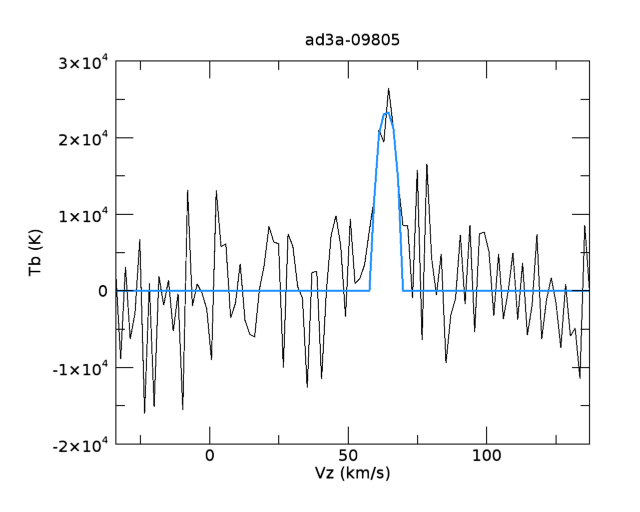}{.33\textwidth}{}}
	\gridline{\fig{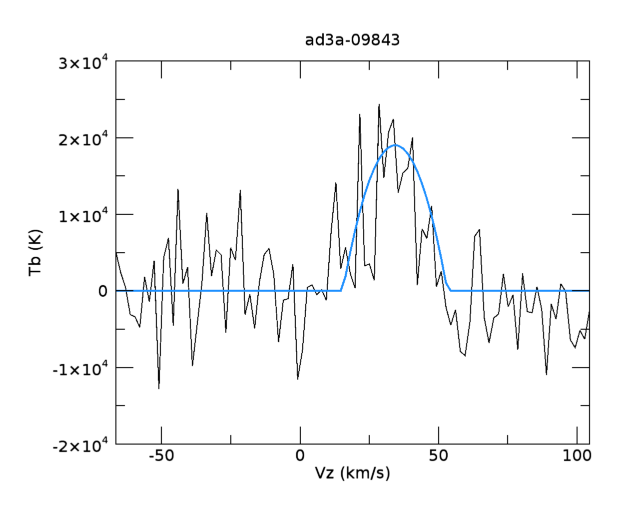}{.33\textwidth}{}
		\fig{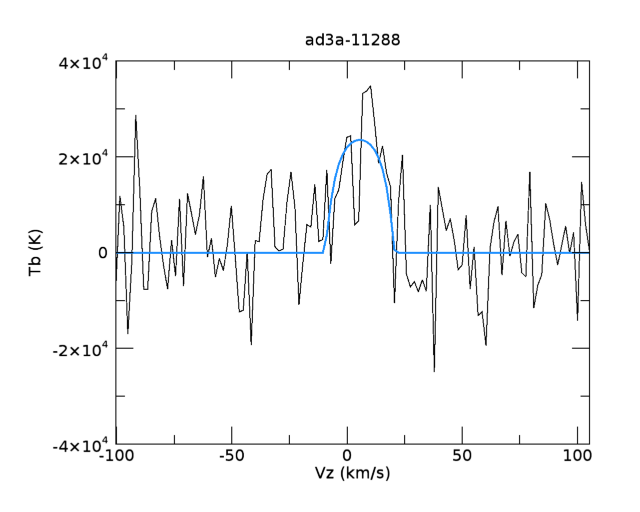}{.33\textwidth}{}
	\fig{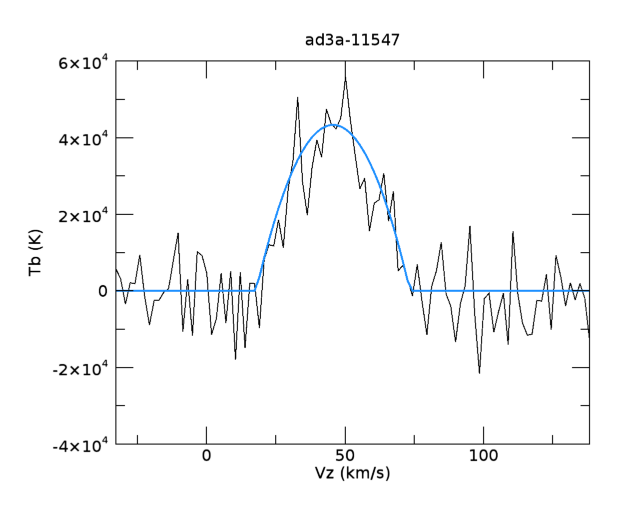}{.33\textwidth}{}}
		\gridline{\fig{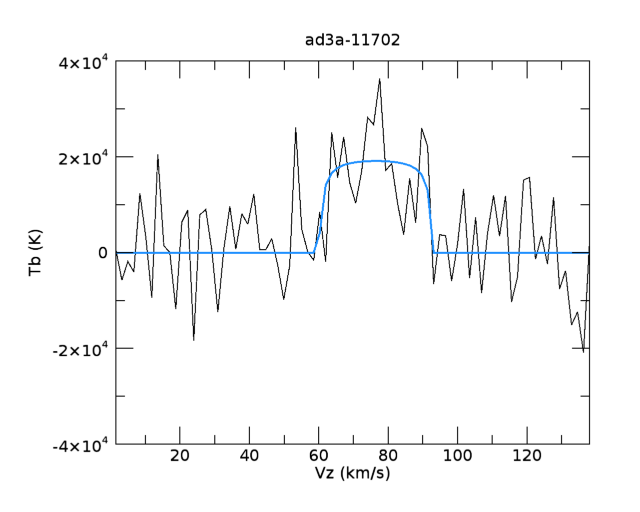}{.33\textwidth}{}
	\fig{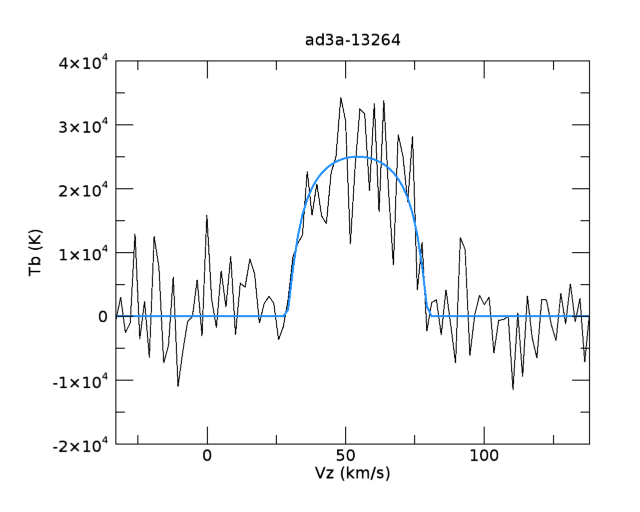}{.33\textwidth}{}
		\fig{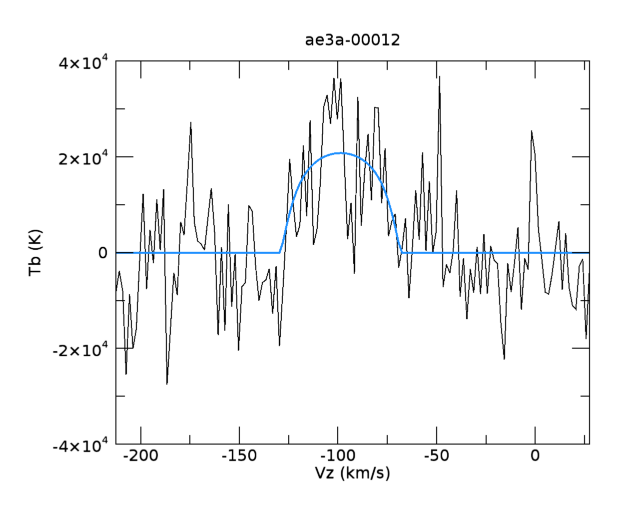}{.33\textwidth}{}}
	\caption{Line profiles of the thermal sources that underwent the fitting process, with
		the blue line indicating the model fit. \label{fittfig}}
\end{figure}

\begin{figure}
	\figurenum{A3}
	\gridline{\fig{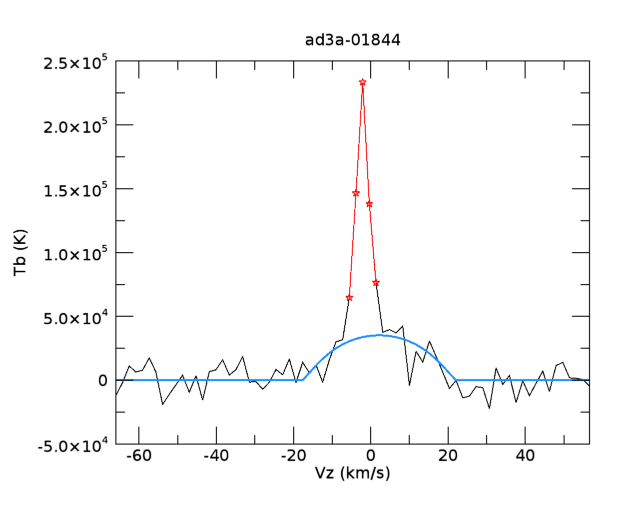}{.33\textwidth}{}
		\fig{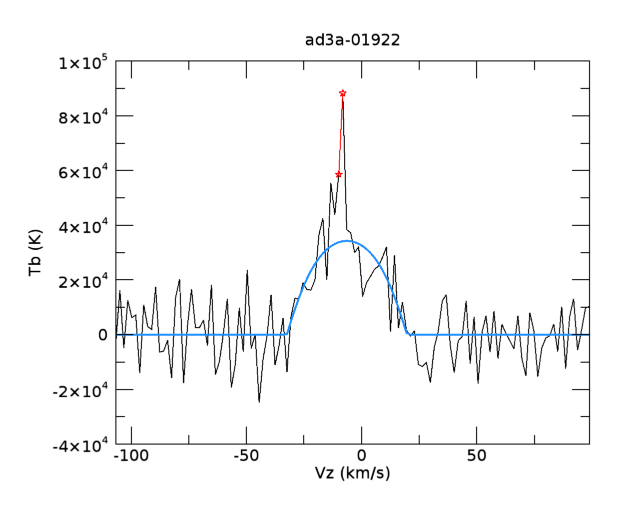}{.33\textwidth}{}
	\fig{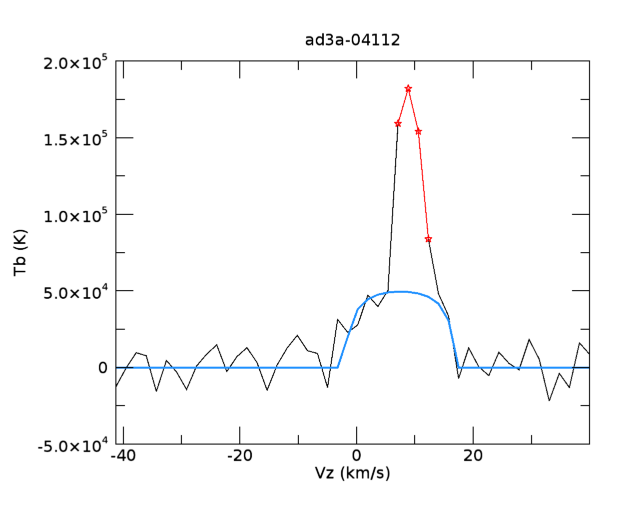}{.33\textwidth}{}}
	\gridline{\fig{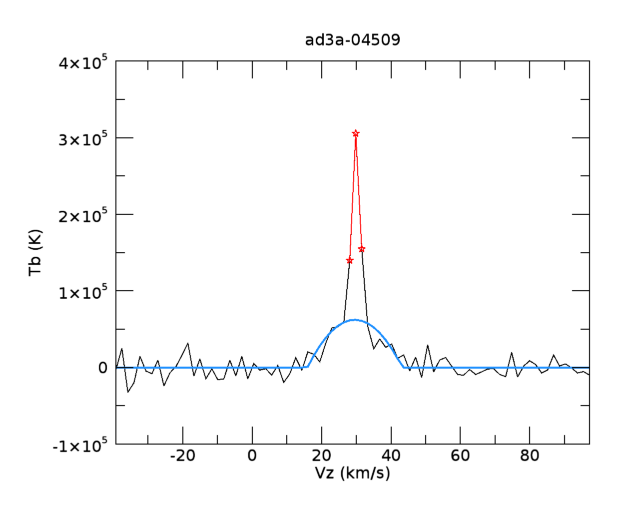}{.33\textwidth}{}
	\fig{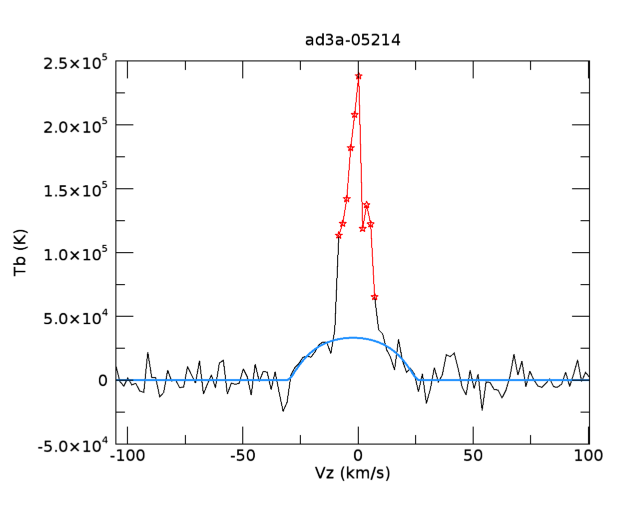}{.33\textwidth}{}
		\fig{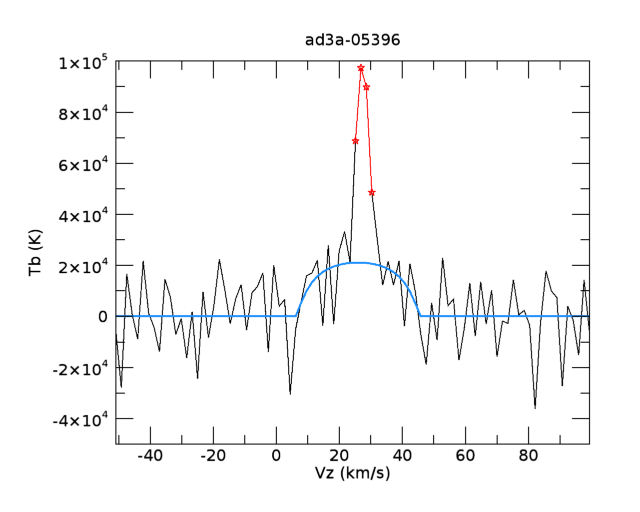}{.33\textwidth}{}}
	\gridline{\fig{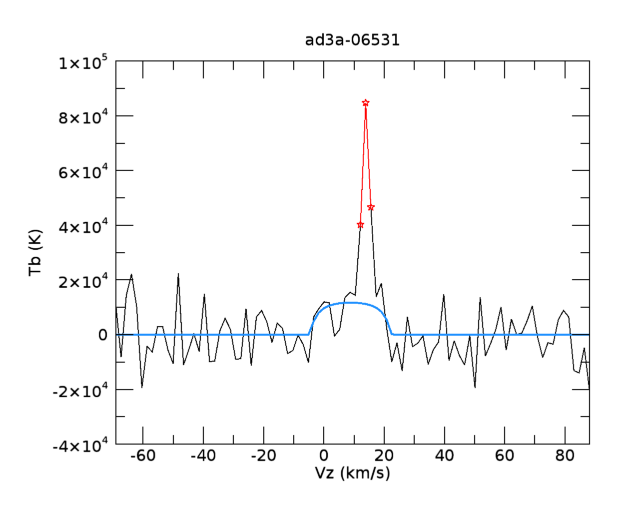}{.33\textwidth}{}
		\fig{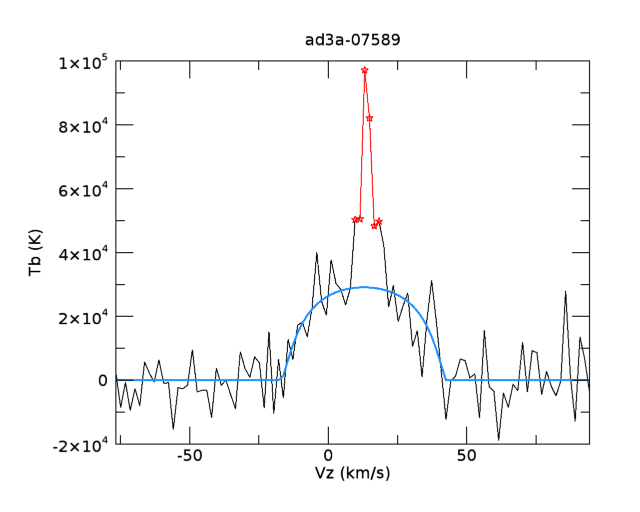}{.33\textwidth}{}
	\fig{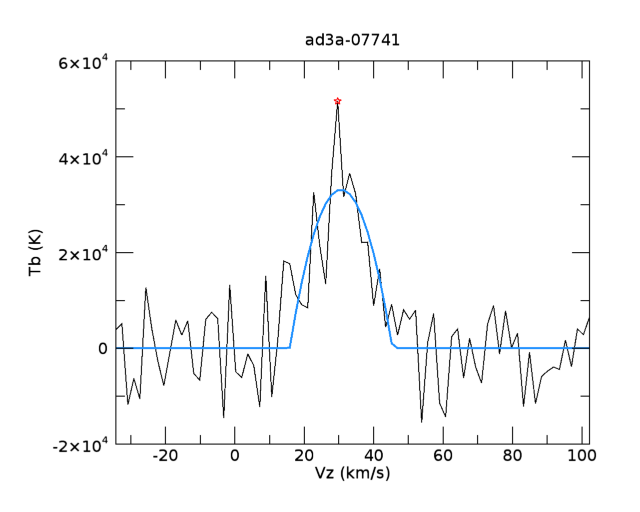}{.33\textwidth}{}}
		\gridline{\fig{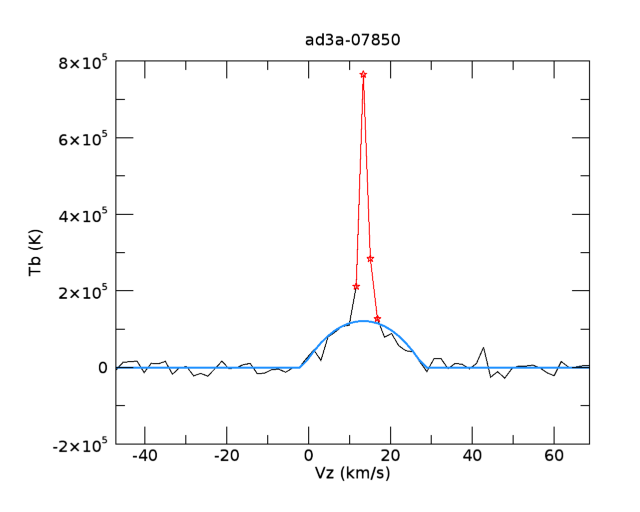}{.33\textwidth}{}
	\fig{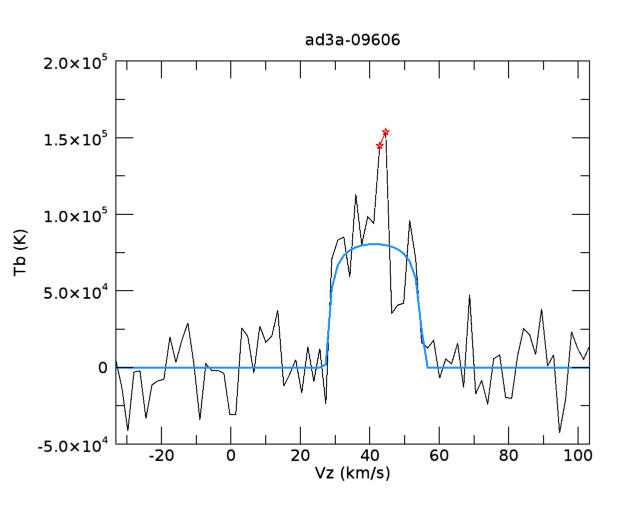}{.33\textwidth}{}
		\fig{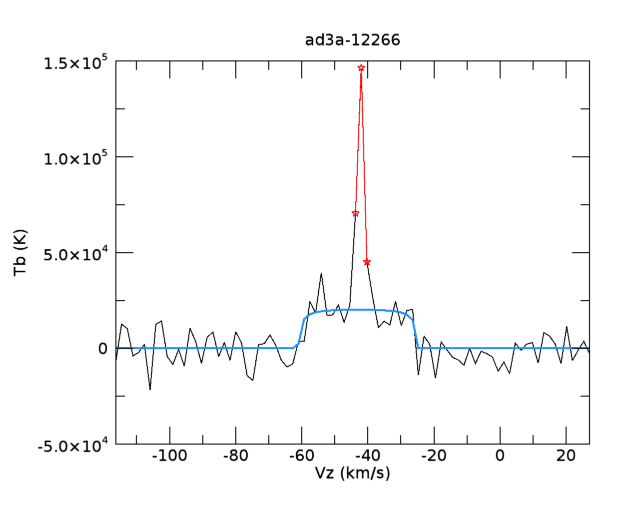}{.33\textwidth}{}}
	\caption{Line profiles of thermal sources with a maser component along with our model fit (overplotted blue line). Maser components were
		zero-weighted for the fit. The zero-weighted points are marked by red stars in these diagrams. \label{fitcfig}}
\end{figure}
\begin{figure}
	\figurenum{A3 (continued)}
	\gridline{\fig{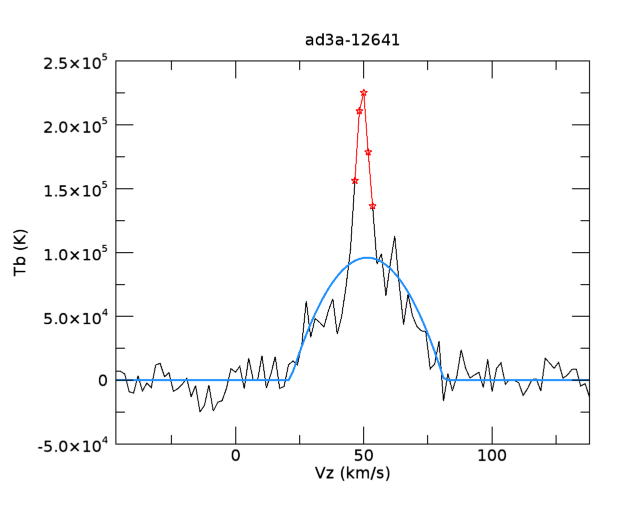}{.33\textwidth}{}
		\fig{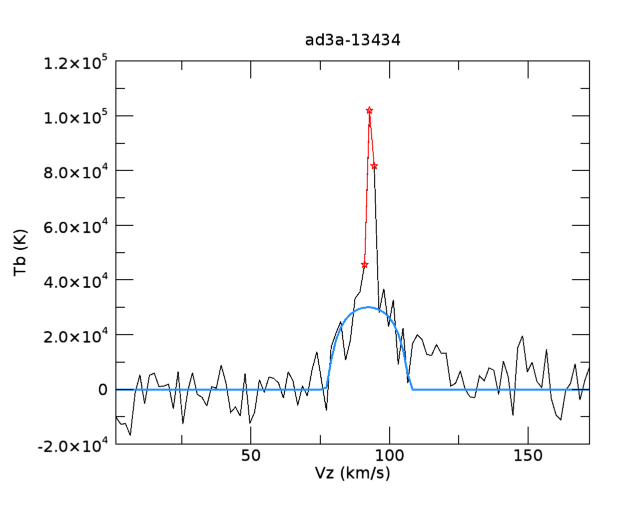}{.33\textwidth}{}
	\fig{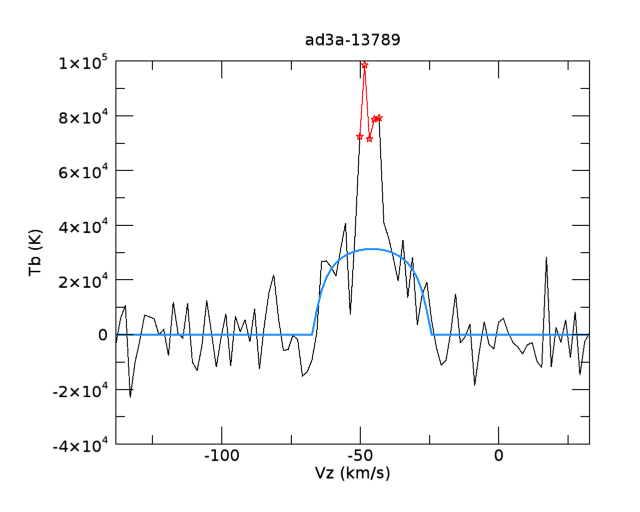}{.33\textwidth}{}}
		\gridline{\fig{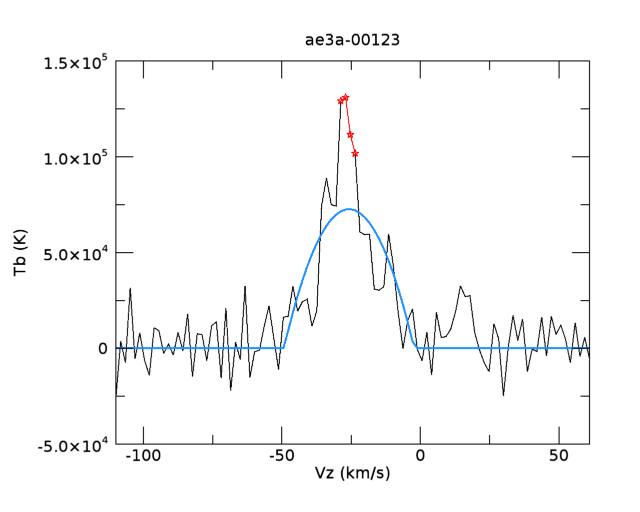}{.33\textwidth}{}
	\fig{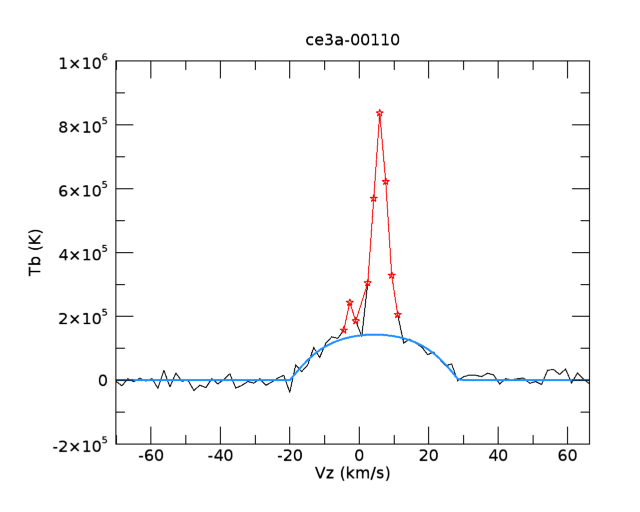}{.33\textwidth}{}
		\fig{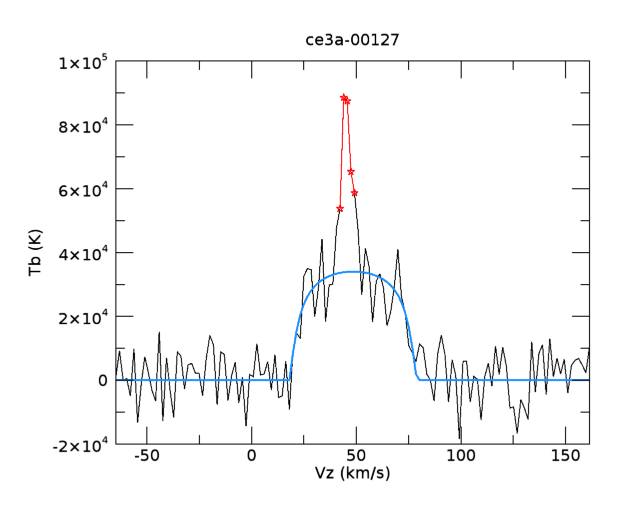}{.33\textwidth}{}}
	\gridline{\fig{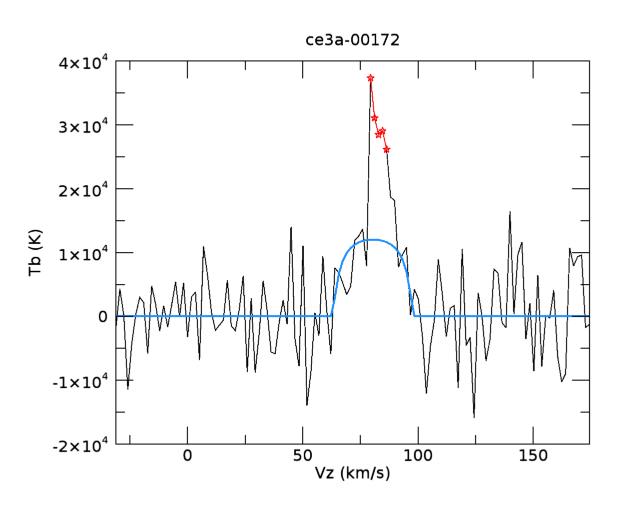}{.33\textwidth}{}
		\fig{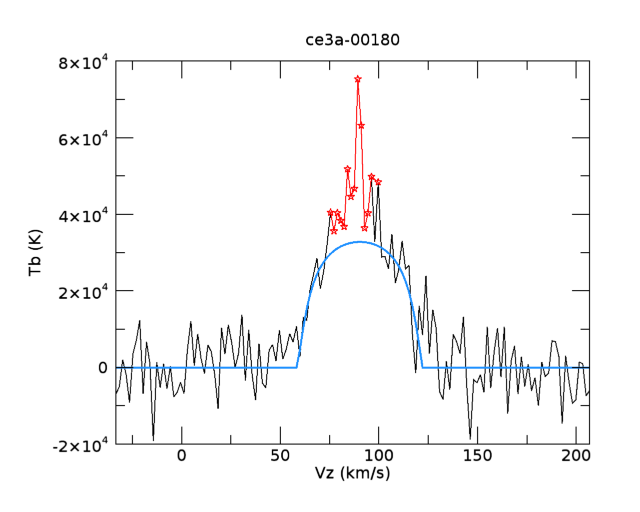}{.33\textwidth}{}}
	\caption{}
\end{figure}

\end{document}